\documentclass[dvipsnames]{lmcs}
\pdfoutput=1

\usepackage[utf8]{inputenc}

\usepackage{lastpage}
\lmcsdoi{17}{3}{2}
\lmcsheading{}{\pageref{LastPage}}{}{}%
{Jul.~05,~2018}{Aug.~10,~2021}{}

\usepackage{verbatim}
\usepackage{doi}
\usepackage{hyperref}
\newcommand\myshade{85}
\colorlet{mylinkcolor}{violet}
\colorlet{mycitecolor}{YellowOrange}
\colorlet{myurlcolor}{Aquamarine}
\hypersetup{ linkcolor  = mylinkcolor!\myshade!black,
             citecolor  = mycitecolor!\myshade!black,
             urlcolor   = myurlcolor!\myshade!black,
             colorlinks = true,
           }
\usepackage{cite} % http://www.ctan.org/pkg/cite
\usepackage{graphicx}
\usepackage{amssymb}
\usepackage{amsmath}
\usepackage{amsthm}
\usepackage{wasysym}
\usepackage{proof}
\usepackage{semantic} % for math ligatures like |=, <->, ->, etc

\newsavebox{\mybox}

\newcommand{\Coloneqq}{\mathrel{::=}}

\newcommand{\VDash}{\mathrel{\mid\!\vDash}}

\usepackage{listings}
\usepackage[all]{xy}
\renewcommand{\vDash}{\models}

\newcommand{\OM}{\ensuremath{\mathcal{O\!M}}}
\newcommand{\FM}{\ensuremath{\mathcal{F\!M}}}
\newcommand{\LM}{\ensuremath{\mathcal{L\!M}}}
\newcommand{\lProlog}{{$\lambda$Prolog}}
\newcommand{\ttt}{\mathtt{t\hspace*{-.25em}t}}
\newcommand{\fff}{\mathtt{f\hspace*{-.25em}f}}

\newcommand{\boxm}[1]{\mathopen{\big[ #1 \big]}} %{[\hspace{.2ex}#1\hspace{.2ex}]}
\newcommand{\diam}[1]{\mathopen{\big\langle #1 \big\rangle}}
\newcommand{\respects}{\mathrel{\textrm{respects}}}%{\ensuremath{\mathop{\scriptstyle{\pmb{\texttt{\char`_}\!\sqcap\!\backslash}\!\medbullet}}}}
\newcommand{\respecting}{\mathrel{\textrm{respecting}}}%{\ensuremath{\mathop{\scriptstyle{\pmb{\texttt{\char`_}\!\sqcap\!\backslash}\!\medbullet}}}}
\newcommand{\cpar}{\mathrel{\mid}} %{\mathrel{\|}}
\newcommand{\co}[1]{\overline{#1}}
\newcommand{\match}[1]{\mathopen{\left[#1\right]}}
\newcommand{\yields}{\supset}
\newcommand{\barb}[2]{#1\!\downarrow\!#2}
\newcommand{\lts}[1]{\xsta{#1}}

\newcommand{\isim}{\mathrel{\sim_{I}}}
\newcommand{\lsim}{\mathrel{\dot\sim_{L}}}
\newcommand{\qsim}{\simeq}

\newcommand{\Late}{\mathrel{\dot{\smash{\vDash}}_{L}}} % the \smash lowers the dot
\newcommand{\xsta}[1]{ 
\setbox0=\hbox{\,${\scriptstyle#1}$\,}
\ifdim\wd0<14pt\wd0=10pt\fi
\mathrel{\raise1.6pt\hbox{$\mathop{\rule{\wd0}{0.62pt}}\limits^{{#1}\;}$}\mkern-6mu{\scriptstyle\blacktriangleright}}
}
\newcommand{\sub}[2]{\mathclose{\{^{#2\!\!\!}\reflectbox{$\!\smallsetminus$}_{\!#1}\hspace{-.2mm}\}}}
\newcommand{\fv}[1]{\mathopen{\mathrm{fn}}\left(#1\right)}
\newcommand{\bn}[1]{\mathopen{\mathrm{bn}}\left(#1\right)}
\newcommand{\fn}[1]{\mathopen{\mathrm{fn}}\left(#1\right)}

\newcommand{\dom}[1]{\mathopen{\mathrm{dom}}\left(#1\right)}
\newcommand{\context}[1]{\mathopen{\mathcal{C}}\left\{\,#1\,\right\}}
\newcommand{\oprivate}[2]{\mathopen{\co{ #1 }}\left( #2 \right)} 
 
\newcommand{\n}[1]{\mathopen{\mathrm{n}}\left(#1\right)}
\newcommand{\bang}{\mathopen{!}}

\def\Ecal{{\mathcal{E}}}

\usepackage[colorinlistoftodos,textsize=tiny]{todonotes}

\title[A Characterisation of Open Bisimilarity]{\texorpdfstring{A
Characterisation of Open Bisimilarity \\ using an Intuitionistic Modal Logic}{A
Characterisation of Open Bisimilarity using an Intuitionistic Modal Logic}}

\author[K.Y.~Ahn]{Ki Yung Ahn\rsuper{a}}
\address{\lsuper{a}Department of Computer Engineering,
Hannam University, Daejeon, Korea}
\email{kya@hnu.kr}

\author[R.~Horne]{Ross Horne\rsuper{b}}
\address{
  \lsuper{b}Computer Science, University of Luxembourg, Esch-sur-Alzette, Luxembourg
}
\email{ross.horne@uni.lu}

\author[A.~Tiu]{Alwen Tiu\rsuper{c}}
\address{\lsuper{c}Research School of Computer Science, Australian National University, Canberra, Australia}
\email{alwen.tiu@anu.edu.au}

% mandatory: Please provide 1-5 keywords

\begin{document}

% make the title area
\maketitle

\begin{abstract}
Open bisimilarity is defined for open process terms in which free variables may appear. The insight is, in order to characterise open bisimilarity, we move to the setting of intuitionistic modal logics. The intuitionistic modal logic introduced, called $\mathcal{OM}$, is such that modalities are closed under substitutions, which induces a property known as \textit{intuitionistic hereditary}. Intuitionistic hereditary reflects in logic the lazy instantiation of free variables performed when checking open bisimilarity. The soundness proof for open bisimilarity with respect to our intuitionistic modal logic is mechanised in Abella. The constructive content of the completeness proof provides an algorithm for generating distinguishing formulae, which  we have implemented. We draw attention to the fact that there is a spectrum of bisimilarity congruences that can be characterised by intuitionistic modal logics.
\end{abstract}

\section{Introduction}\label{sec:intro}

This work provides insight into the logical nature of open bisimilarity~\cite{sangiorgi96acta}, but firstly we recall why open bisimilarity itself is important.
An asset of open bisimilarity is that it defines a congruence relation for open process terms, i.e., process terms containing free variables.
Recall that the original notions of bisimilarity proposed for the $\pi$-calculus (early and late bisimilarity~\cite{Milner92,Milner93}) do not directly define congruence relations for open process terms.
Having a bisimilarity that is a congruence for open process terms improves compositional reasoning, since, having established an algebraic property, we can apply the property with confidence, anywhere inside a process, even under constructs such as input prefixes that bind variables. 
By providing a notion of bisimilarity that is a congruence for open process terms, open bisimilarity provides a method for the $\pi$-calculus that stays true to this desirable property of a processes algebra.

Besides improved algebraic properties, open bisimilarity can be used to improve the efficiency of equivalence checking.
For example, open bisimilarity is the notion of bisimilarity implemented in the Mobility Workbench~\cite{Victor1994} -- the first toolkit for the $\pi$-calculus;
and in the tool SPEC~\cite{Tiu2016} -- an equivalence checker for the spi-calculus, useful for verifying cryptographic protocols.
A reason open bisimilarity is efficient to implement is that it allows a lazy approach to instantiating variables.
When we perform an input action, we are not required to explore all possible inputs. Instead, we can represent the input value as a variable symbolically representing all possible inputs.
This symbolic approach to inputs can avoid unnecessarily exploring hyper-exponentially many inputs; instead, exploring only the state space necessary.
This lazy ``call-by-need'' approach to input transitions is particularly useful when checking bisimilarity for applied extensions of the $\pi$-calculus, where infinitely many messages may be received for a single input action~\cite{briais06entcs,Tiu2010,Horne2020}. Thus open bisimilarity has impact beyond the setting of the $\pi$-calculus.

The trick for ensuring that open bisimilarity is a congruence, and also for permitting a lazy approach to inputs is as follows: an open bisimulation is closed under all permitted substitutions at every step in the bisimulation game.
When we move to the setting of logic, closure under substitutions corresponds to a concept called \textit{intuitionistic hereditary}, which can be used to induce an intuitionistic logic~\cite{Kripke1965}.
This observation leads us to the intuitionistic modal logic in this work.

To understand why closing under substitutions results in an intuitionistic modal logic, firstly consider the setting of a \textbf{classical modal logic}.
In a classical setting, the law of excluded middle holds, hence we expect that  $\diam{\tau}\ttt \vee \neg\diam{\tau}\ttt$ is a tautology.
That is, any process can either perform a $\tau$-transition or it cannot perform a $\tau$-transition.

In contrast, now consider the setting of an \textbf{intuitionistic modal logic}.
In the intuitionistic setting we close under all substitutions, so $P \vDash \neg\diam{\tau}\ttt$ now reads, \textit{under any substitution} $\sigma$, process $P\sigma$ cannot perform a $\tau$-transition.
Under this interpretation we have the following.
\[
 \co{a}b \cpar c(x) \not\vDash \neg\diam{\tau}\ttt
\]
To see why the above is not satisfiable, observe that, by applying substitution $\sub{a}{c}$ to the above process, we reach process $\co{c}b \cpar c(x)$, which is a $\pi$-calculus process for which a communication is enabled on the channel represented by variable $c$. Since we have demonstrated there is a substitution under which a $\tau$-transition can be performed, process $\co{a}b \cpar c(x)$ cannot satisfy formula $\neg\diam{\tau}\ttt$ in the intuitionistic setting.

As in the classical case, in the intuitionistic case we have the following, since there is a substitution under which no communication can be performed (the identity substitution).
\[
 \co{a}b \cpar c(x) \not\vDash \diam{\tau}\ttt
\]
Putting the above together, we have the following in our intuitionistic modal logic, since we have just shown that neither branch of the disjunction is satisfiable.
\[
 \co{a}b \cpar c(x) \not\vDash \diam{\tau}\ttt \vee \neg\diam{\tau}\ttt
\]
Notice the above we claimed was a tautology in the classical case, since it is an instance of the law of excluded middle.
Hence the above example demonstrates that, by closing operators of the modal logic under substitutions, the law of excluded middle does not hold.
The absence of the law of excluded middle is a key criterion for any intuitionistic logic.

Intuitively, the absence of the law of excluded middle for the example above can be interpreted as follows.
For $\co{a}b \cpar c(x)$, we have not yet decided whether the process can perform a $\tau$-transition or not perform a $\tau$-transition.
It is possible that $a$ and $c$ could be the same channel but, since they are variables, we have not yet decided whether this is the case.

So, inducing the key feature of an open bisimulation, closure under substitutions, in a modal logic gives rise to an intuitionistic modal logic.
Furthermore, we establish in this work that such an intuitionistic modal logic, called $\mathcal{OM}$, characterises open bisimilarity.
In the tradition pioneered by Hennessy and Milner~\cite{HenMil85},
a modal logic characterises a bisimilarity whenever, given two processes, they are bisimilar if and only if there is no {\em distinguishing formula} separating them.
A distinguishing formula is a formula that holds for one process but does not hold for the other process.
Such distinguishing formulae are useful for explaining why two processes are not bisimilar, since when processes are not bisimilar we can always exhibit a distinguishing formula.

As an example of a distinguishing formula, consider the following two processes.
\[
R \triangleq
\mathopen{\tau.}\left(\co{a}{b}.{{a}{(x)}} + {a}{(x)}.{\co{a}{b}} + \tau\right)
   +
\mathopen{\tau.}\left(\co{a}{b}.{{c}{(x)}} + {c}{(x)}.{\co{a}{b}}\right)
\qquad\qquad
S \triangleq R + \mathopen{\tau.}\left(\co{a}{b} \cpar {c}{(x)}\right)
\]
\noindent
The above processes are not open bisimilar.  
Process $R$ satisfies $\mathopen{\boxm{\tau}}\left( \diam{\tau}\ttt \vee \neg \diam{\tau}\ttt \right)$, where the box modality indicates that
$\diam{\tau}\ttt \vee \neg \diam{\tau}\ttt$ holds for {\em{all}} processes reachable by applying a substitution and then a $\tau$-transition.
However, process $S$ does not satisfy $\mathopen{\boxm{\tau}}\left( \diam{\tau}\ttt \vee \neg \diam{\tau}\ttt \right)$, since there is a $\tau$-transition to process $\co{a}b \cpar c(x)$
that we just agreed does not satisfy $\diam{\tau}\ttt \vee \neg \diam{\tau}\ttt$.
In this example, the absence of the law of excluded middle is necessary in order for a formula distinguishing these processes to exist in $\OM$.

Modal logics characterising late bisimilarity and early bisimilarity were developed early in the literature on the $\pi$-calculus, by Milner, Parrow and Walker~\cite{Milner93}, as part of the motivation for the $\pi$-calculus itself.
However, proving that a modal logic can characterise open bisimilarity was an open problem until a solution was provided in the conference version of this paper~\cite{Ahn2017}.
This extended version includes more details on proofs, new examples, details on the mechanisation of soundness, and further insight into the spectrum of bisimilarity congruences that can be characterised by variants of our intuitionistic modal logic. We also show that soundness and completeness results for $\OM$ extend from finite processes to infinite but finitely branching processes, without changing the logic, since finite distinguishing strategies are sufficient to distinguish such processes.

A key novelty of this work is the constructive proof of completeness of this logical characterisation.
Due to the intuitionistic nature of the modal logic, the completeness proof cannot appeal to certain classical principles, such as de Morgan dualities.
This forces the proof to follow a strategy quite different to corresponding completeness proofs for classical modal logics. The proof directly constructs a pair of distinguishing formulae for every pair of processes that are not open bisimilar.

\subsection*{Outline}
Section~\ref{sec:direct} introduces the semantics of intuitionistic modal logic \OM.
Section~\ref{sec:openbisim} recalls open bisimilarity and states the soundness and completeness results.
Section~\ref{sec:complete} presents the proof of the correctness of an algorithm for generating distinguishing formulae, which is used to establish completeness of $\OM$ with respect to open bisimilarity.
Section~\ref{sec:related} situates $\OM$\ with respect to other modal logics in the spectrum of classical and intuitionistic notions of bisimilarity, highlighting that open bisimilarity is not a canonical notion of bisimilarity congruence for the $\pi$-calculus and that other bisimilarity congruences can also be characterised by intuitionistic modal logics, including a new notion of late bisimilarity congruence, called intermediate bisimilarity, introduced for this discussion.
Section~\ref{sec:abella} describes how the proof assistant Abella~\cite{Baelde14} was used to mechanically prove soundness of \OM\ with respect to open bisimulation.
The soundness theorem (Section~\ref{sec:abella}) and selected examples (Section~\ref{sec:direct} and Section~\ref{sec:complete}) have been mechanised in the Abella theorem prover, and are 
available online.\footnote{Via \url{https://github.com/alwentiu/abella/tree/master/pic}}
Section~\ref{sec:dfalgo} demonstrates an implementation of the algorithm automatically generating distinguishing formulae, extracted from the proof of Proposition~\ref{proposition:non-bisim}.

\section{Introducing the intuitionistic modal logic \OM}
\label{sec:direct}

\begin{figure}
{\small
\[
\begin{tabular}{l@{\hspace{1ex}}r@{\hspace{1ex}}l@{\hspace{1ex}}r}
$\pi$&$\Coloneqq$&$\tau$ & (progress) \\
&& $\co{x}{z}$ & (free output) \\
&& $\oprivate{x}{z}$ & (bound output) \\
&& $x(z)$ & (input)\\
\\\\
$P$&$\Coloneqq$ & $0$ & (deadlock)\\
&& $\nu x. P$ & (new)\\
&& $\pi.P$ & (action)\\
&& $\match{x=y}P$ & \quad(match) \\
&& $P \cpar P$ & (par) \\
&& $P + P$ & (choice) \\
&& $\bang P$ & (replication)
\end{tabular}
\qquad\quad
\begin{gathered}
\infer{
\pi.P \lts{\pi} P
}{}
\qquad\qquad
\infer[\text{$x \not\in \n{\pi}$}]{
\nu x. P \lts{\pi} \nu x . Q
}{
P \lts{\pi} Q
}
\\[5pt]
\infer[x \not= z]{
\nu z. P \lts{\oprivate{x}{z}} Q
}{
P \lts{\co{x}z} Q
}
\qquad
\infer[
\begin{smallmatrix}
\text{if $x \in \bn{\pi}$} 
\\
\text{then $x$ fresh for $R$} 
\end{smallmatrix}
]{
P \cpar R \lts{\pi} Q \cpar R
}{
P \lts{\pi} Q
}
\\[5pt]
\infer{
P + Q \lts{\pi} R
}{
P \lts{\pi} R
}
\qquad\quad
\infer{
[x=x]P \lts{\pi} R
}{
P \lts{\pi} R
}
\qquad\quad
\infer{
\bang P
\lts{\pi}
Q
}{
P
\lts{\pi}
Q
}
\\[5pt]
\infer{
P \cpar Q
\lts{\tau}
\mathopen{\nu z .}\left(P' \cpar Q'\right)
}{
P \lts{\oprivate{x}{z}} P'
\quad
Q \lts{x(z)} Q'
}
\qquad
\infer{
P \cpar Q
\lts{\tau}
P' \cpar Q'\sub{z}{y}
}{
P \lts{\co{x}{y}} P'
\quad
Q \lts{x(z)} Q'
}
\\[5pt]
\infer{
\bang P
\lts{\tau}
\mathopen{\nu z .}\left(P' \cpar Q'\right)
}{
P \lts{\oprivate{x}{z}} P'
\quad
P \lts{x(z)} Q'
}
\qquad\quad
\infer{
\bang P
\lts{\tau}
P' \cpar Q'\sub{z}{y}
}{
P \lts{\co{x}{y}} P'
\quad
P \lts{x(z)} Q'
}
\end{gathered}
\]
}
\caption{Syntax and semantics of the $\pi$-calculus, omitting symmetric rules for choice and parallel composition,
where $\n{x(y)} = \n{\oprivate{x}{y}} = \n{\co{x}y} = \left\{x,y\right\}$, \;
$\bn{x(y)} = \bn{\oprivate{x}{y}} = \left\{ y \right\}$, $\n{\tau} = \bn{\tau} = \bn{\co{x}{y}} = \emptyset$,
and $\fn{\pi} = \n{\pi}\setminus\bn{\pi}$.
Processes $\nu x . P$, $z(x).P$ and $\bar{z}(x).P$ bind $x$ in $P$.
}
\label{figure:pi}
\end{figure}

We recall the syntax and labelled transition system of the $\pi$-calculus (Fig.~\ref{figure:pi}).
Note all the atomic symbols $x, y, \hdots$ are variables.
There is no separate syntactic class for channels or names in this presentation of the $\pi$-calculus.
Distinctions between the roles of variables
are made by the use of binders: variables may appear as open variables, be bound by an input binder, or by a new name binder.
Notably, the new name binder $\nu x.P$ indicates any occurrence of $x$ in $P$ is a ground name that is distinct from any other ground name and cannot be guessed by an observer
unless it is provided explicitly to the observer through an output action\footnote{An explanation, coming from computational security, is the new name binder represents a nonce selected from a set of nonces that is exponential in size w.r.t.\ some sufficiently large parameter, hence cannot be guessed by an observer.}.
Variables may also be bound by an input binder, say $z(x).P$ where occurrences of variable $x$ in $P$ are treated as placeholders for some message (also represented by a variable) that will be received when an input on a channel represented by variable $z$ occurs. Variables that are not bound, i.e., free variables, 
are critical for this call-by-need approach to the $\pi$-calculus
where they are used as symbolic placeholders that range over all possible ways in which they may be instantiated.

Other features include:
the deadlocked process that can do nothing, output prefixes that output a free variable or extrude a variable bound by a new name binder on a channel, 
the silent progress action $\tau$,
the match guard that tests for equality, 
parallel composition, and non-deterministic choice.
We also include a replication operator, which creates unboundedly many parallel copies of a process.

Transitions are labelled with four types of action ranged over by $\pi$: free outputs, bound outputs, inputs and internal progress ($\tau$). 
A \textit{free output} represents sending a free variable, whereas a \textit{bound output} represents extruding a bounded name.
We employ a late labelled transition system for the $\pi$-calculus, where the 
variable on the \textit{input} action is a symbolic placeholder that need not be instantiated until after an input transition.
The action $\tau$ represents some internal communication, resulting from the synchronisation of an input and output action.
We use the notations $\bn{E}$ and $\fv{E}$ to represent the bound variables and, respectively, free variables
in a given expression $E$ (processes, actions, formulae, etc.). 
We assume $\alpha$-conversion for bound variables. 

Histories are used in the definitions of both the intuitionistic modal logic and open bisimilarity.
Histories are lists representing what is known about free variables due to how they have been communicated previously to the environment.
There are two types of information about variables recorded in a history.
Variables $x$, that were bound by a new name binder and have been \textit{extruded} using output action $\co{a}(x)$, we call \textit{private names}, and denote them in histories by $x^o$.
Variables $z$, symbolically representing the possible messages received by an input action $a(z)$, are denoted in histories by $z^i$.
What matters is the alternation in the history between variables representing extruded private names and variables representing symbolic inputs:
if an input variable is to be instantiated with a private name, the private name must have been extruded by an earlier output in the history.
This is reflected in the following definition of a respectful substitution.
\begin{defi}[$\sigma \respects h$]\label{def:respects}
A history is a (dot separated) list of variables annotated with $o$ or $i$.
Substitution $\sigma$ \textit{respects} history $h$ whenever,
for all $h'$ and $h''$ such that $h = h' \cdot x^o \cdot h''$,
$x\sigma = x$, and for all $y \in \fn{h'}$, we have $y\sigma \not= x$.
Here $\fn{h'}$ is all the variables appearing anywhere in $h'$.
\end{defi}
For example, substitution $\sub{z}{y}$ respects history $x^i\cdot y^o\cdot z^i$, since input variable $z$ appears after $y$ was output, hence $y$ was known to the environment at the time $z$ was input.
In contrast, substitution $\sub{x}{y}$ does not respect history $x^i\cdot y^o\cdot z^i$, since variable $x$ was input before private name $y$ was output.

\begin{rem}
Note that histories fulfil the role of sets of inequality constraints called \textit{distinctions} in the original work on open bisimilarity~\cite{sangiorgi96acta}.
Although distinctions are more general than histories, it is shown in~\cite{TiuMil10} that given a history $h$ and its corresponding distinction $D$, the corresponding definitions of open bisimilarity coincide.

Histories effectively form a symbolic constraint system restricting the use of variables. It is worth noting that the effect of histories can also be achieved by maintaining a set of fresh name constraints, indicating that a private name output is fresh with respect to the free variables in the process at the moment the private name was output, as proposed for symbolic approaches to the $\psi$-calculus~\cite{Johansson2012}. In order to capture open bisimilarity using such a symbolic constraint systems, care needs to be taken to ensure that the constraint system is interpreted intuitionistically.
\end{rem}

\subsection{The semantics of the intuitionistic modal logic \OM}
\label{sec:om}

The syntax for modal logic \OM\ extends intuitionistic logic with equality and modalities, as follows.

\begin{defi} 
The syntax for modal logic \OM\ is defined by the following grammar.
\[
\begin{array}[t]{lr}
\left.
\begin{array}{rlr}
\phi \Coloneqq & \ttt & \mbox{top} \\
      \mid& \fff & \mbox{bottom} \\
      \mid& \phi \wedge \phi & \mbox{and} \\
      \mid& \phi \vee \phi & \mbox{or} \\
      \mid& \phi \yields \phi & \qquad\mbox{implies} \\
      \mid& x = y & \mbox{equality} 
\end{array}\qquad
\right\}
\qquad \mbox{intuitionistic logic}
\\\qquad\!\!
\left.
  \hspace{2.9pt}\begin{array}{rlr}
  \mid& \diam{\pi}\phi & \qquad\!\mbox{diamond} \\[2.6pt]
      \mid& \boxm{\pi}\phi & \mbox{box}
\end{array}
\qquad
\right\}
\qquad \mbox{modalities}
\end{array}
\]
Formulae of the form $\boxm{\co{a}(x)}\phi$, $\boxm{{a}(x)}\phi$, $\diam{\co{a}(x)}\phi$, or $\diam{{a}(x)}\phi$, bind variable $x$ in $\phi$.
\label{def:OMsyntax}
\end{defi}

\begin{figure}
\[
\begin{array}{lcl}
P \vDash^{h} \ttt  &  & \mbox{always holds.} \\
P \vDash^{h} \fff  &  & \mbox{never holds.} \\
P \vDash^{h} x=y   & \mbox{iff} & \mbox{$x$ and $y$ are the same variable.} \\
P \vDash^{h} \phi_1 \land \phi_2 &\mbox{iff}&
  P \vDash^{h} \phi_1 ~\mbox{and}~ P \vDash^{h} \phi_2.
\\
P \vDash^{h} \phi_1 \lor \phi_1 &\mbox{iff}&
  P \vDash^{h} \phi_1 ~\mbox{or}~ P \vDash^{h} \phi_2.
\\
P \vDash^{h} \phi_1 \yields \phi_1 &\mbox{iff}&
 \forall \sigma \respecting h,
  P\sigma \vDash^{h\sigma} \phi_1\sigma ~ \mbox{ implies } ~ P\sigma \vDash^{h\sigma} \phi_2\sigma.
\\
P \vDash^{h} \diam{\alpha}\phi &\mbox{iff}&
  \exists\,Q,~ P \lts{\alpha} Q ~\mbox{and}~ Q \vDash^{h} \phi.
\\
P \vDash^{h} \diam{\co{a}(z)}\phi &\mbox{iff}&
  \exists\,Q,~ P \lts{\co{a}(z)} Q ~\mbox{and}~ Q \vDash^{h\cdot z^o} \phi.
\\
P \vDash^{h} \diam{{a}(z)}\phi &\mbox{iff}&
  \exists\,Q,~ P \lts{{a}(z)} Q ~\mbox{and}~ Q \vDash^{h\cdot z^i} \phi.
  \\[2.6pt]
P \vDash^{h} \boxm{\alpha}\phi &\mbox{iff}&
  \forall \sigma \respecting h,
    \forall Q,\, P\sigma \lts{\alpha\sigma} Q  \; \mbox{ implies } \;
    Q \vDash^{h\sigma} \phi\sigma.
\\
P \vDash^{h} \boxm{\co{a}(z)}\phi &\mbox{iff}&
  \forall\sigma \respecting h,~ \forall Q, P\sigma \lts{\co{a\sigma}(z)} Q \mbox{ implies } Q \vDash^{h\sigma \cdot z^o} \phi\sigma.
\\
 P \vDash^{h} \boxm{{a}(z)}\phi &\mbox{iff}&
  \forall \sigma \respecting h, \forall Q, 
    P\sigma \lts{a\sigma(z)} Q \mbox{ implies } Q \vDash^{h\sigma \cdot z^i} \phi\sigma.
\end{array}
%\end{array}
\]
In each of the above, $\alpha$ is of the form $\tau$ or $\co{a}b$; and $z$ is fresh for $h$, and $\sigma$.
\caption{Semantics of the modal logic \OM.}
\label{figure:om}
\end{figure}

The semantics of intuitionistic modal logic \OM, presented in Fig.~\ref{figure:om}, is defined in terms of the late labelled transition system in Fig.~\ref{figure:pi} and history respecting substitutions (Definition~\ref{def:respects}).
Satisfaction is defined as follows, by treating all free variables as inputs in the past.
\begin{defi}[satisfaction]\label{def:modal}
Satisfaction, written $P \vDash \phi$, holds whenever
$P \models^{x_0^i \cdot \hdots \cdot x_n^i} \phi$ according to the inductive definition in Fig.~\ref{figure:om},
where $\fv{P} \cup \fv{\phi} \subseteq \left\{x_0, \hdots, x_n\right\}$.
Note $\fv{P}$ and $\fv{\phi}$ are simply the free variables in $P$ and $\phi$ respectively.
\end{defi}

Both modalities are closed under all respectful substitutions.
However, observe in Fig.~\ref{figure:om} there is an asymmetry in the definition of these modalities.
In contrast to the box modality, the definition of the diamond modality need not be closed under all respectful substitutions.

To explain this asymmetry between the box and diamond modalities in the definitions, observe, 
for the diamond modality, a transition must be possible regardless of the substitution. 
Thus it is sufficient to check the identity substitution.
For example, the following is not satisfiable.
\[
  \match{x = y}\tau \not\vDash \diam{\tau}\ttt
\]
To check the above does not hold, it is sufficient to check that $\match{x = y}\tau$ cannot perform a $\tau$-transition.
This is reflected in the semantics of the diamond modalities.

In contrast, for the box modality there may exist substitutions other than the identity substitution enabling a transition,
hence we should consider all respectful substitutions. 
Perhaps the simplest example, requiring closure of box under respectful substitutions, is the following:
\[
 \match{x = y}\tau \vDash \boxm{\tau}\left( x = y \right)
\]
The above satisfaction holds since for any substitution $\sigma$ such that $\left(\match{x = y}\tau\right)\mathclose{\sigma} \lts{\tau} 0$ it must be case that $x\sigma = y\sigma$.
Thus for all such substitutions we have $0 \vDash  x\sigma = y\sigma$ holds, as required.
In contrast, observe the above process does not satisfy $\boxm{\tau}\fff$.

Intuitionistic hereditary establishes that all formulae are closed under respectful substitutions.
We state this property of $\OM$ as a lemma, since it will be used in the completeness proof later in this paper.
\begin{lem}[intuitionistic hereditary]\label{lemma:mono}
If $P \vDash^h \phi$ holds then $P\theta \vDash^{h\theta} \phi\theta$ holds
for any $\theta$ respecting $h$.
\end{lem}
The hereditary lemma suggests a Kripke model for $\OM$ that satisfies the usual frame conditions for intuitionistic logic: consider respectful substitutions as a binary relation between `worlds', where a world is simple a set of equalities between variables.  Then it can be proven that  this gives rise to an intuitionistic Kripke frame. The satisfaction relation in Figure~\ref{figure:om} can be reformulated using this notion of worlds explicitly. The interested reader may consult Appendix~\ref{sec:kripke} for details of a Kripke semantics for $\OM$. We note that the Kripke semantics presented there is not needed to prove the main results of this paper; hence can be safely skipped.

\subsection{Checking the law of excluded middle is invalidated.}\label{sec:middle}
Given the semantics in Fig.~\ref{figure:om}, we can now formally check examples from the introduction.
We claimed $
\co{a}b \cpar c(x) \not \vDash \diam{\tau}\ttt
\vee \neg \diam{\tau}\ttt
$, where $\neg \phi$, as standard, is defined as $\phi \yields \fff$.
This example demonstrates the law of excluded middle is invalid.
Appealing to the rule for disjunction, observe that we have the following.
\[
\co{a}b \cpar c(x) \not\vDash \diam{\tau}\ttt
\qquad
\mbox{and}
\qquad
\co{a}b \cpar c(x) \not\vDash  \neg \diam{\tau}\ttt
\]
The former can hold only if $\co{a}b \cpar c(x)$ is guaranteed to make a $\tau$-transition;
but such a transition is only possible under a substitution $\sigma$ such that $a\sigma=c\sigma$,
hence $\co{a}b \cpar c(x) \not\vDash \diam{\tau}\ttt$.
For the latter, 
we should consider all substitutions 
which enable a $\tau$-transition;
and, since such a substitution $\sub{a}{c}$ exists,
$\co{a}b \cpar c(x) \not\vDash \neg \diam{\tau}\ttt$.

Critically for this work, the above example illustrates that a property typically used to establish the completeness of open bisimilarity with respect to 
a classical modal logic breaks down.
In the \textbf{classical} setting, we expect $P \not\vDash \phi$ if and only if $P \vDash \neg\phi$.
However, as the above example demonstrates, there are processes, such as $\co{a}b \cpar c(x)$, that do not satisfy $\diam{\tau}\ttt$, but also do not satisfy $\neg\diam{\tau}\ttt$.
Hence in the intuitionistic setting we \textbf{cannot} appeal to this principle of classical modal logic.

As a further example of this principle, observe the following are both unsatisfiable.
\[
\tau \not\vDash \boxm{\tau}\left(x = y\right)
\qquad
\mbox{and}
\qquad
\tau \not\vDash  \neg \boxm{\tau}\left( x = y \right)
\]
The former is unsatisfiable since, under the identity substitution, $\tau \lts{\tau} 0$, but $0 \not\vDash x = y$.
The latter is also unsatisfiable since, there is a substitution $\sub{x}{y}$ such that $\tau\sub{x}{y} \lts{\tau} 0$ still holds and $0 \vDash x\sub{x}{y} = y\sub{x}{y}$ holds;
but clearly $0 \vDash \fff\sub{x}{y}$ can never hold; hence $\tau \not\vDash \neg \boxm{\tau}\left( x = y \right)$.

As expected for an intuitionistic logic, further classical dualities break, as witnessed by the following examples of unsatisfiable formulae.
\[
\match{x = y}\tau \not\vDash \neg\neg\diam{\tau}\ttt \yields \diam{\tau}\ttt
\qquad
\mbox{and}
\qquad
0 \not\vDash \neg\neg\mathopen{\neg}\left(x = y\right) \yields \mathopen{\neg}\left(x = y\right)
\]
Also de Morgan dualities cannot be applied. For example, \textbf{classically} we have $P \vDash \mathopen{\neg}\left( \boxm{\tau}\fff \wedge \diam{\tau}\ttt \right)$ if and only if $P \vDash \diam{\tau}\ttt
\vee \boxm{\tau}\fff$.
However in the \textbf{intuitionistic} setting we have the following.
\[
\co{a}b \cpar c(x) \vDash \mathopen{\neg}\left( \boxm{\tau}\fff
\wedge \diam{\tau}\ttt \right)
\qquad
\mbox{but}
\qquad
\co{a}b \cpar c(x) \not \vDash \diam{\tau}\ttt
\vee \boxm{\tau}\fff
\]
Note, in this paper, intuitionistic negation is used only to explain such examples illustrating the intuitionistic nature of \OM. Results in subsequent sections do not depend on intuitionistic negation. However, related work~\cite{Horne2018} highlights that intuitionistic negation has a role when logically characterising open bisimilarity for processes with mismatch (inequality guards, which can model the else branch of an \texttt{if-then-else} statement). Thus this formulation of $\OM$ is robust for some useful extensions of the $\pi$-calculus.

\section{Open bisimilarity, soundness and completeness}
\label{sec:openbisim}
We recall the definition of open bisimilarity. Open bisimilarity is the greatest symmetric relation
closed under all respectful substitutions and labelled transitions at every step.
Notice we use the history to record whenever a (symbolic) input or private output occurs.
\begin{defi}[open bisimilarity]\label{def:bisim}
An open bisimulation $\mathcal{R}$ is a symmetric relation on processes, indexed by a history $h$,
such that: if $P \mathrel{\mathcal{R}^h} Q$ then, the following hold:
\begin{itemize}
\item For all substitutions $\sigma$ respecting $h$, we have $P\sigma \mathrel{\mathcal{R}^{h\sigma}} Q\sigma$.
\item If $P \lts{\alpha} P'$, then there exists $Q'$ such that $Q \lts{\alpha} Q'$ and $P' \mathrel{\mathcal{R}^{h}} Q'$, where $\alpha$ is a $\tau$ or $\co{a}b$.
\item If $P \lts{\co{a}(x)} P'$, for $x$ fresh for $h$ and $Q$, there exists $Q'$ such that $Q \lts{\co{a}(x)} Q'$ and $P' \mathrel{\mathcal{R}^{h\cdot x^o}} Q'$.
\item If $P \lts{{a}(x)} P'$, for $x$ is fresh for $h$, there exists $Q'$ such that $Q \lts{a(x)} Q'$ and $P' \mathrel{\mathcal{R}^{h\cdot x^i}} Q'$.
\end{itemize}
Open bisimilarity, written $P \sim Q$, is defined whenever there exists
an open bisimulation $\mathcal{R}$ such that $P \mathrel{\mathcal{R}^{x_0^i \cdot \hdots \cdot x_n^i}} Q$,
where $\fv{P} \cup \fv{Q} \subseteq \left\{x_0, \hdots, x_n\right\}$.
\end{defi}

The main result of this paper is that, for finite $\pi$-calculus processes
open bisimilarity is characterised by $\mathcal{OM}$ formulae.
This result is broken into soundness and completeness of the intuitionistic modal logic characterisation.
\begin{thm}[soundness]\label{theorem:sound}
If $P \sim Q$ then for all $\mathcal{OM}$ formulae $\phi$, $P \vDash \phi$ iff $Q \vDash \phi$.
\end{thm}
\begin{thm}[completeness]\label{theorem:complete} 
If we have that 
for all $\mathcal{OM}$ formulae $\phi$, $P \vDash \phi$ iff $Q \vDash \phi$,
then $P \sim Q$.
\end{thm}

The proof of soundness has been mechanically checked in the proof assistant Abella \cite{Baelde14} using the two-level logic approach~\cite{GacMilNad12} to reason about the $\pi$-calculus semantics specified in \lProlog~\cite{NadMil88}.
The proof of soundness proceeds by induction on the structure of formulae.
An explanation of the soundness proof and mechanisation we defer until Section~\ref{sec:abella}.

The proof of completeness is explained in detail in Section~\ref{sec:complete}.
Before providing proofs, we provide examples demonstrating the implications of Theorems~\ref{theorem:sound} and~\ref{theorem:complete}.
Due to soundness, if two processes are bisimilar, we cannot find a distinguishing formula that holds for one process but does not hold for the other process.
Due to completeness, if two process are not open bisimilar, then we can construct a distinguishing formula that holds for one process but does not hold for the other process.
Thus \OM\ formulae can be used as a certificate that can be presented as efficiently checkable evidence to explain why two processes are not open bisimilar.

\subsection{Sketch of algorithm for generating distinguishing formulae}
\label{sec:sketch}
The completeness proof, explained later in Section~\ref{sec:complete}, contains an algorithm for
generating distinguishing formulae for processes that are not open bisimilar.
Here, we provide a sketch of the algorithm executed on key examples.

\subsubsection{Example requiring intuitionistic assumptions}\label{eg:law}
The algorithm proceeds over the structure of a tree of moves that show two processes are not open bisimilar -- the \textit{distinguishing strategy}.
In the base case, we have a pair of processes where, under a substitution, one process can make a transition, but the other process cannot match the transition.
We provide two examples of applying the base case to obtain formulae.

\begin{description}
  \item[$\match{x=y}\tau \not\sim \tau$] 
    The distinguishing strategy for these processes is as follows: the process $\tau$ leads with transition $\tau \lts{\tau} 0$, but $\match{x\theta=y\theta}\tau$ can make
    a $\tau$-transition only when $x\theta = y\theta$.
    From this distinguishing strategy we generate two formulae, one biased to each process\footnote{By ``biased'', we mean the formula is satisfied by the process indicated but not by the other process considered.}.
    Since process $\tau$ leads in the distinguishing strategy, $\diam{\tau}\ttt$ is a distinguishing formula biased to process $\tau$, as follows.
\[
\tau \vDash \diam{\tau}\ttt
\qquad
\mbox{and}
\qquad
\match{x=y}\tau \not\vDash \diam{\tau}\ttt
\]
As remarked in the previous section, negating formula $\diam{\tau}\ttt$ does not provide a formula biased to $\match{x=y}\tau$.
To construct a formula biased towards $\match{x=y}\tau$, write down a box modality $\boxm{\tau}$ followed by the strongest postcondition that holds after a $\tau$-transition is enabled, i.e. $x = y$. 
This gives rise to the following distinguishing formula, as required.
\[
\match{x=y}\tau \vDash \boxm{\tau}\left(x = y\right)
\qquad
\mbox{and}
\qquad
\tau \not\vDash \boxm{\tau}\left(x = y\right)
\]

\item[$\match{x=y}\tau \not\sim 0$] 
For these processes the distinguishing strategy is $\left(\match{x=y}\tau\right)\sub{x}{y} \lts{\tau} 0$, but $0$ cannot make
    a $\tau$-transition, under any substitution.
To construct a distinguishing formula biased to $\match{x=y}\tau$, we write down $x = y$ as the weakest pre-condition under which a $\tau$-transition is enabled, expressed as follows.
\[
\match{x=y}\tau \vDash \left(x=y\right) \yields \diam{\tau}\ttt
\qquad
\mbox{and}
\qquad
0 \not\vDash \left(x=y\right) \yields \diam{\tau}\ttt
\]
To construct a formula biased to $0$, write $\boxm{\tau}$ followed by $\fff$, which, vacuously, is the strongest postcondition guaranteed after $0$ performs a $\tau$-transition, since no $\tau$-transition is enabled under any substitution. This gives us the following distinguishing formula.
\[
0 \vDash \boxm{\tau}\fff
\qquad
\mbox{and}
\qquad
\match{x=y}\tau \not\vDash \boxm{\tau}\fff
\]
\end{description}

Now consider the \textit{inductive case} of an algorithm for constructing distinguishing formulae.
In the inductive cases, two processes cannot be distinguished by an immediate transition.
However, under some substitution, 
one process can make a $\pi$ transition to a state, say $P'$, but,
under the same substitution the other process can only make a corresponding $\pi$ transition
to reach states $Q_i$ that are not open bisimilar to $P'$.
This allows a distinguishing formula to be inductively constructed from the distinguishing formulae for $P'$
paired with each $Q_i$.

For example, consider how to construct distinguishing formulae for processes $P$ and $Q$ below.
\begin{quote}
$\!\!\!\!
\xymatrix@1@=7pt@M=2pt{
P \ar[d]^{\tau}
  & \!\!\!\!\triangleq \tau.\match{x=y}\tau + \tau + \tau.\tau &
   \qquad \nsim \qquad&
  \tau + \tau.\tau \triangleq \!\!\!\!\! & Q
  \ar[d]_{\tau}
  \ar@/^/[drr]^{\tau}
  \\
\match{x=y}\tau& \qquad \qquad ~~
    & & & 0 
    & & \tau &
} $
\vspace{-.6ex}
\end{quote}

Observe from the above transitions, that the process $P$ can perform a $\tau$-transition to a state $\match{x =y}\tau$ 
that is not bisimilar to any state reachable by a $\tau$-transition from process $Q$.
Process $Q$ may perform $\tau$-transitions either to $\tau$ or $0$.
However we have just seen above that $\match{x=y}\tau \not\sim 0$ and $\match{x=y}\tau \not\sim \tau$;
hence we have a distinguishing strategy.

The distinguishing strategies and distinguishing formulae for the above base cases, enable us to construct distinguishing formulae for this inductive case.
The distinguishing formula satisfied by $P$ is a \textit{diamond} modality followed by
the \textit{conjunction} of the distinguishing formulae biased to $\match{x=y}\tau$ in each base case above, as follows.
\[
P
\vDash
\mathopen{\diam{\tau}}
\left(
 \boxm{\tau}\left( x=y \right) \wedge \left( x=y \yields \diam{\tau}\ttt \right)
\right)
\quad
\mbox{and}
\quad
Q
\not\vDash
\mathopen{\diam{\tau}}
\left(
 \boxm{\tau}\left( x=y \right) \wedge \left( x = y \yields \diam{\tau}\ttt \right)
\right)
\]
The distinguishing formula satisfied by $Q$ is a \textit{box} followed by
the \textit{disjunction} of the formulae not satisfied by $\match{x=y}\tau$ in each of the base cases above, as follows:
\[
P \not\vDash 
\mathopen{\boxm{\tau}}\left(\diam{\tau}\ttt \vee \boxm{\tau}\fff\right)
\qquad
\mbox{and}
\qquad
Q
\vDash 
\mathopen{\boxm{\tau}}\left(\diam{\tau}\ttt \vee \boxm{\tau}\fff\right)
\]
To confirm that the above are indeed distinguishing formulae for $P$ and $Q$, assume for contradiction that 
$
Q
\vDash
\mathopen{\diam{\tau}}\left(
 \boxm{\tau}\left( x = y \right)
 \wedge
 \left( x = y \yields \diam{\tau}\ttt\right)
\right)
$
holds.
By definition of diamond modalities, this holds iff
either $0 \vDash \boxm{\tau}\left( x = y \right) \wedge \left( x = y \yields \diam{\tau}\ttt\right)$
or $\tau \vDash \boxm{\tau}\left( x = y \right)  \wedge \left( x = y \yields \diam{\tau}\ttt\right)$ holds.
Observe that
$0 \vDash x = y \yields \diam{\tau}\ttt$ holds
iff we make the additional assumption that $x$ and $y$ are persistently distinct, i.e., we have additional assumption $\mathopen{\neg}\left( x = y \right)$.
In addition, observe that $\tau \vDash \boxm{\tau}\left( x = y \right)$ holds iff we make the additional assumption that $x = y$.

Indeed, by these observations, we know that the following hold: 
\[
\tau + \tau.\tau 
\vDash
\left(x = y \vee \mathopen{\neg}\left( x = y \right)\right)
\yields
\mathopen{\diam{\tau}}\left(
 \boxm{\tau}\left( x = y \right)
 \wedge
 \left( x = y \yields \diam{\tau}\ttt\right)
\right)
\]
\[
\tau + \tau.\tau 
\vDash
\mathopen{\diam{\tau}}\left(
 \boxm{\tau}\left( x = y \right)
 \wedge
 \left( x = y \yields \diam{\tau}\ttt\right)
\right)
\yields
\left(x = y \vee \mathopen{\neg}\left( x = y \right)\right)
\]
Notice that $x = y \vee \mathopen{\neg}\left( x = y \right)$ is an instance of the law of excluded middle for equality;
hence, in the \textbf{classical} setting, assuming the law of excluded middle, the formula above biased to $Q$ is also satisfied by $P$;
and vice versa.
Indeed there would be no distinguishing formulae for processes $P$ and $Q$;
and hence in a classical framework the modal logic would be incomplete for open bisimilarity.

Similarly, in the \textbf{intuitionistic} setting, we can mechanically prove the following.
\[
\tau + \tau.\tau + \tau.\match{x = y}\tau
\vDash
\left(x = y \vee \mathopen{\neg}\left( x = y \right)\right)
\yields
\mathopen{\boxm{\tau}}\left(\diam{\tau}\ttt \vee \boxm{\tau}\fff\right)
\]
\[
\tau + \tau.\tau + \tau.\match{x = y}\tau
\vDash
\mathopen{\boxm{\tau}}\left(\diam{\tau}\ttt \vee \boxm{\tau}\fff\right)
\yields
\left(x = y \vee \mathopen{\neg}\left( x = y \right)\right)
\]
Since intuitionistic logics do not assume the law of excluded middle, as long as we evaluate the semantics of $\mathcal{OM}$ in an intuitionistic framework, we have distinguishing formulae.
We have formalised in Abella the above four examples of satisfaction involving the law of excluded middle.

\subsubsection{Example involving private names that are distinguishable}\label{eg:privdist}
Respectful substitutions ensure that a private name can never be input earlier than it was output.
Consider the following processes.
\[
P\triangleq \nu x.\co{a}x.a(y).\tau
\quad
\nsim
\quad
\nu x.\co{a}x.a(y).\match{x=y}\tau
\triangleq Q
\]

These processes are not open bisimilar because $P$ can make the following three transition steps:
$\nu x.\co{a}x.a(y).\tau \lts{\co{a}(x)} a(y).\tau \lts{a(y)} \tau \lts{\tau} 0$.
However, $Q$ can only match the first two steps. At the third step, a base case of the distinguishing formula algorithm for $\tau \not\sim^{a^i\cdot x^o\cdot y^i} \match{x=y}\tau$ applies.
In this case, any substitution $\theta$ respecting $a^i\cdot x^o\cdot y^i$
enabling transition $\match{x=y}\tau\theta \lts{\tau} 0$ is such that $y\theta = x$ and $x\theta = x$; hence $x\theta = y\theta$.
Hence we have the following formulae biased to each process.
\[
\match{x=y}\tau \vDash^{a^i\cdot x^o\cdot y^i} \boxm{\tau}\left( x = y\right)
\qquad
\mbox{and}
\qquad
\tau \vDash^{a^i\cdot x^o\cdot y^i} \diam{\tau}\ttt
\]
By applying inductive cases of the distinguishing formulae algorithm to the input and output actions, we obtain the following two distinguishing formulae.
\[
\nu x.\co{a}x.a(y).\tau \vDash \diam{\co{a}(x)}\diam{a(y)}\diam{\tau}\ttt
\quad
\mbox{and}
\quad
\nu x.\co{a}x.a(y).\match{x=y}\tau \vDash \boxm{\co{a}(x)}\boxm{a(y)}\boxm{\tau}\left( x = y \right)
\]

\subsubsection{Example involving private names that are indistinguishable}\label{eg:privnondist}
In contrast to the previous example, consider the following processes
where the process on the right extrudes a private name and then compares it to a free variable.
\[
\nu x.\co{a}x \quad\sim\quad \nu x.\co{a}x.\match{x=a}\tau
\]
These processes are open bisimilar, hence by Theorem~\ref{theorem:sound}
there is no distinguishing formula. The existence of a distinguishing formula
of the form $\diam{\co{a}(x)} \left(x=a \yields \diam{\tau}\ttt\right)$ is \emph{prevented} by the history.
For example, both $\nu x.\co{a}x.\match{x=a}\tau \vDash \diam{\co{a}(x)}\left(x=a \yields \diam{\tau}\ttt\right)$
and $\nu x.\co{a}x  \vDash \diam{\co{a}(x)}\left(x=a \yields \diam{\tau}\ttt\right)$ hold.

To see why, observe $\nu x.\co{a}x \vDash^{a^i} \diam{\co{a}(x)}\left(x=a \yields \diam{\tau}\ttt\right)$
holds
if and only if
$\nu x.\co{a}x \lts{\co{a}(x)} 0$
and $0 \vDash^{a^i\cdot x^o} x=a \yields \diam{\tau}\ttt$.
By definition of implication, 
this holds if only if, for all $\theta$ respecting $a^i\cdot x^o$ and such that $x\theta = a\theta$,
we have $0 \vDash^{a^i\cdot x^o} \diam{\tau}\ttt$.
However, there is no substitution $\theta$ respecting $a^i \cdot x^o$ such that $x\theta = a\theta$.
By the definition of a respectful substitution, 
$\theta$ must satisfy $x\theta = x$ and $x \not= a\theta$, contradicting constraint $x\theta = a\theta$.
Thereby $0 \vDash^{a^i\cdot x^o} x=a \yields \diam{\tau}\ttt$ holds vacuously;
hence we have that $\nu x.\co{a}x  \vDash^{a^i} \diam{\co{a}(x)} \left(x=a \yields \diam{\tau}\ttt\right)$ holds as required.

\section{Completeness of open bisimilarity with respect to \OM}
\label{sec:complete}

In order to prove completeness we first provide a direct definition of what it means for two processes to be not open bisimilar,
which we refer to as \textit{distinguishability}.
Since open bisimilarity is defined in terms of a greatest fixed point of relations satisfying a certain closure property,
distinguishability is defined in terms of a least fixed point satisfying the dual property.
This leads to the direct definition of distinguishability in this section.

Since distinguishability is defined in terms of a least fixed point, there is a distinguishing strategy, consisting of a finite tree of moves.
We inductively define distinguishability in terms of a family of relations on processes indexed by a history $\not\sim_n$, for $n \in \mathbb{N}$.
The base case is when, for some respectful substitution one player can make a move, that move cannot be matched by the other player without applying an additional substitution.
We then define inductively, the family of relations $P \not\sim_n^h Q$ containing all processes that can be distinguished by a strategy with depth at most $n$, i.e.,\ at most $n$ moves are required to reach a pair of processes distinguished according to relation $\not\sim_{0}$, at which point, as just explained above, there is a process reachable by a respectful substitution that can make a move that the other process cannot match under the same substitution.

\begin{defi}[distinguishability]\label{def:non-bisim}
The relation $\not\sim_0$ is the least 
relation, indexed by a history, such that $P \not\sim^h_0 Q$ holds whenever there exist action $\pi$ and substitution $\sigma$ respecting $h$ such that
one of the following holds:
\begin{itemize}
\item
there exists process $P'$ such that $P\sigma \lts{\pi\sigma} P'$ and there is no $Q'$ such that $Q\sigma \lts{\pi\sigma} Q'$, or
\item there exist process $Q'$ such that
$Q\sigma \lts{\pi\sigma} Q'$ and there is no $P'$ such that $P\sigma \lts{\pi\sigma} P'$.
\end{itemize}
In both cases, we require that if $x \in \bn{\pi}$, then $x$ is fresh for $P\sigma$, $Q\sigma$ and $h\sigma$.

Inductively, $\not\sim_{n+1}$ is the least 
relation extending $\not\sim_{n}$ such that 
$P \not\sim^h_{n+1} Q$ whenever 
for some substitution $\sigma$ respecting $h$, one of the following holds, where, in the following, $\alpha$ is $\tau$ or $\co{a}b$
and $x$ is fresh for $P\sigma$, $Q\sigma$ and $h\sigma$:
\begin{itemize}
\item $\exists P'.~P\sigma \lts{\alpha\sigma} P'$ and $\forall Q_i$ such that $Q\sigma \lts{\alpha\sigma} Q_i$, $P' \not\sim^{h\sigma}_n Q_i$, or
\item $\exists P'.~P\sigma \lts{\co{a\sigma}(x)} P'$, and, $\forall Q_i$ such that $Q\sigma \lts{\co{a\sigma}(x)} Q_i$, $P' \not\sim^{h\sigma \cdot x^o}_n Q_i$, or
\item $\exists P'.~P\sigma \lts{a\sigma(x)} P'$, and, $\forall Q_i$ such that $Q\sigma \lts{a\sigma(x)} Q_i$, $P' \not\sim^{h\sigma \cdot x^i}_n Q_i$, or
\item $\exists Q'.~Q\sigma \lts{\alpha\sigma} Q'$ and $\forall P_i$ such that $P\sigma \lts{\alpha\sigma} P_i$, $P_i \not\sim^{h\sigma}_n Q'$, or
\item $\exists Q'.~Q\sigma \lts{\co{a\sigma}(x)} Q'$, and,  $\forall P_i$ such that $P\sigma \lts{\co{a\sigma}(x)} P_i$, $Q' \not\sim^{h\sigma \cdot x^o}_n P_i$, or
\item $\exists Q'.~Q\sigma \lts{a\sigma(x)} Q'$, and,  $\forall P_i$ such that $P\sigma \lts{a\sigma(x)} P_i$, $Q' \not\sim^{h\sigma \cdot x^i}_n P_i$.
\end{itemize}

The relation $\not\sim$, pronounced distinguishability, is defined to be the least relation containing $\not\sim_n$ for all $n \in \mathbb{N}$,
i.e.\ $\bigcup_{n\in\mathbb{N}} \not\sim_n$.
Define $P \not\sim Q$ whenever $P \not\sim^{x_1^i\cdot\hdots x_m^i} Q$ where $\fv{P}\cup\fv{Q} \subseteq \left\{ x_1, \hdots x_m \right\}$.
\end{defi}

It is immediate from the definition that distinguishability is symmetric.
\begin{lem}\label{lemma:symmetry}
The relations $\not\sim$ and $\not\sim_n$, for all $n \geq 0$, are symmetric.
\end{lem}

It is an established result that, for the version of the $\pi$-calculus with replication that we employ, image finiteness holds. 
\begin{lem}[image finiteness~\cite{Sangiorgi1995}]\label{lemma:finite}
For process $P$ and action $\pi$
there are 
finitely many $P_i$, up to $\alpha$-conversion, such that $P \lts{\pi} P_i$.
\end{lem}

Distinguishability in Definition~\ref{def:non-bisim}
coincides with the negation of open bisimilarity in Definition~\ref{def:bisim}, which relies on the fact that we consider an image-finite process calculus in this work.
\begin{lem}\label{lemma:not}
$P \not\sim Q$ (Def.~\ref{def:non-bisim}) does not hold,
if and only if $P \sim Q$ (Def.~\ref{def:bisim}) holds.
\end{lem}
\begin{proof}
In order to establish the forward implication, 
we construct a relation
such that $P \mathrel{\mathcal{R}^{h}} Q$ whenever  $\fv{P} \cup \fv{Q}  \subseteq \fv{h}$ 
and there does not exists $n$ such that $P \not\sim_n^{h} Q$.
We then show that $\mathcal{R}$ is an open bisimulation.
Symmetry of $\mathcal{R}$ is immediate from Lemma~\ref{lemma:symmetry}.
Below we consider the remaining 
cases required to show that $\mathcal{R}$ is an open bisimulation.

\begin{description}[leftmargin=2.5mm]
\item[{Case of a respectful substitution}]
Assume that $P \mathrel{\mathcal{R}^{h}} Q$ holds and 
$\theta$ respects $h$,
and suppose for contradiction that $P\theta \mathrel{\mathcal{R}^{h\theta}} Q\theta$ does not hold.
Thus there exists $n$ such that $P\theta \mathrel{\not\sim_n^{h\theta}} Q\theta$.
In the case $n = 0$,
there exists $\sigma$ respecting $h\theta$ and action $\pi\theta$
 such that
either: $P\theta\sigma \lts{\pi\theta\sigma} P'$
but there is no $Q'$ such that $Q\theta\sigma \lts{\pi\theta\sigma} Q'$;
or  $Q\theta\sigma \lts{\pi\theta\sigma} Q'$
but there is no $P'$ such that $P\theta\sigma \lts{\pi\theta\sigma} P'$.
Without loss of generality, consider the former case,
where $P\theta\sigma \lts{\pi\theta\sigma} P'$
but there is no $Q'$ such that $Q\theta\sigma \lts{\pi\theta\sigma} Q'$,
and observe that, since $\theta \cdot \sigma$ respects $h$,
by definition, we have $P \not\sim_0^h Q$ holds, contradicting the assumption that $P \mathrel{\mathcal{R}^{h}} Q$ holds.
A similar argument yields a contradiction in the case that $n > 0$.
Therefore $P\theta \mathrel{\mathcal{R}^{h\theta}} Q\theta$ holds.

\item[{Case of a free transition}]
Assume that $P \mathrel{\mathcal{R}^{h}} Q$ and 
$P \lts{\alpha} P'$ hold, where $\alpha = \tau$ or $\alpha = \co{x}z$.
For contradiction, suppose there is no $Q'$ such that $Q \lts{\alpha} Q'$.
Thereby $P \not\sim^h_0 Q$, contradicting the assumption that $P \mathrel{\mathcal{R}^{h}} Q$;
hence there is at least one $Q_i$ such that $Q \lts{\alpha} Q_i$.
By image finiteness, there are finitely many such $Q_i$ (quotienting by $\alpha$-conversion).
Now, for contradiction, assume that $P' \mathrel{\mathcal{R}^{h}} Q_i$
does not hold for all $i$. Hence for all $i$ there exists $n_i$
such that $P' \mathrel{\not\sim_{n_i}^{h}} Q_i$.
Now let $n = \max_i\left\{ n_i \right\}$,
and observe that for all $i$, $P' \mathrel{\not\sim_{n}^{h}} Q_i$,
thereby $P \mathrel{\not\sim_{n+1}^{h}} Q_i$, contradicting the assumption that $P \mathrel{\mathcal{R}^{h}} Q$.
Thus for some $i$, we have $P' \mathrel{\mathcal{R}^{h}} Q_i$, as required.

\item[{Case of a bound output transition}]
Assume that $P \mathrel{\mathcal{R}^{h}} Q$ and 
$P \lts{\co{x}(z)} P'$ hold, where $z$ is fresh for $P$, $Q$ and $h$.
For contradiction, suppose there is no $Q'$ such that $Q \lts{\co{x}(z)} Q'$.
Thereby $P \not\sim_0^h Q$, contradicting the assumption that $P \mathrel{\mathcal{R}^{h}} Q$;
hence for some $Q_i$ we have $Q \lts{\co{x}(z)} Q_i$.
By image finiteness, there are finitely many such $Q_i$ (quotienting by $\alpha$-conversion).
Now assume that for all $i$, we have $P' \mathrel{\mathcal{R}^{h\cdot x^o}} Q_i$
does not hold.
Hence for all $i$, there exists $n_i$ such that $P' \mathrel{\not\sim_{n_i}^{h\cdot x^o}} Q_i$.
Now let $n = \max_i\left\{ n_i \right\}$,
and observe that for all $i$, $P' \mathrel{\not\sim_{n}^{h\cdot x^o}} Q_i$,
hence $P \mathrel{\not\sim_{n+1}^{h}} Q_i$, contradicting the assumption that $P \mathrel{\mathcal{R}^{h}} Q$.
Thus for some $i$, we have $P' \mathrel{\mathcal{R}^{h\cdot x^o}} Q_i$, as required.

\item[{Case of an input transition}] This is almost identical to the case for bound output transitions.
\end{description}
Thus $\mathcal{R}$ is an open bisimulation.
Now assume $P \not\sim Q$ does not hold, 
hence we have $P \mathrel{\mathcal{R}^{x_1^i\cdot\hdots x_m^i}} Q$
holds, where $\fv{P}\cup\fv{Q} \subseteq \left\{ x_1, \hdots x_m \right\}$;
thereby, by definition of open bisimilarity, $P \sim Q$ holds, as required.
The converse direction is immediate from the definitions.
\end{proof}

Notice that distinguishability in Def.~\ref{def:non-bisim} requires that a distinguishing strategy is finite, and finite distinguishing strategies are sufficient to distinguish processes whose transition systems are image finite.
In a more general setting where we do not have image finiteness (such as for weak variants of open bisimilarity or where infinitely branching process 
are permitted in an extended process language) such a finite notion of distinguishability would not suffice.
In such a setting without image finiteness, in order for the negation of bisimilarity and distinguishability to coincide, distinguishability must be defined in terms of transfinite induction.
Hence we would be required to extend $\OM$ with features offering additional distinguishing power, for example with least and greatest fixed points as in the $\mu$-calculus~\cite{Kozen83}, or with finitely supported conjunctions as in related work on nominal transition systems~\cite{Parrow19}.
Here we stick to the setting where we have image finiteness, hence, by Lemma~\ref{lemma:not}, we can rely on finite distinguishing strategies. An investigation of the more general setting where we cannot rely on image finiteness is proposed as future work.

\subsection{Preliminaries}
For the completeness proof that follows, we require the following terminology for substitutions,
and abbreviations for formulae. These are mainly standard.
\begin{defi}\label{def:defs}
Composition of substitutions $\sigma$ and $\theta$ is defined such that $x \left(\sigma \cdot \theta\right) = \left(x \sigma\right) \theta$, for all $x$.
For substitutions $\sigma$ and $\theta$,
$\sigma \leq \theta$ holds whenever there exists $\sigma'$ such that $\sigma \cdot \sigma' = \theta$.
	For a finite substitution $\sigma = {\sub{x_1}{z_1}\cdots\,\sub{x_n}{z_n}}$ formula $\boxm{\sigma}\phi$ abbreviates formula $\left(x_n=z_n\right) \yields \hdots \left(x_1=z_1\right) \yields \phi$.
For finite set of formulae $\phi_i$, formula $\bigvee_i \phi_i$ abbreviates $\phi_1 \vee \!\hdots\! \vee \phi_n$, where the empty disjunction is $\fff$.
Similarly $\bigwedge_i \phi_i$ abbreviates $\phi_1 \wedge \hdots \wedge \phi_n$, where the empty conjunction is $\ttt$.
\end{defi}

We require the following technical lemmas.
The first unfolds the definition of $\boxm{\sigma}\phi$ in Def.~\ref{def:defs} above.
The second is required in inductive cases involving bound output and input.
The third is a monotonicity property for transitions, along with side conditions for the bookkeeping of bound names that may appear in labels.
\begin{lem}
\label{lemma:box}
If for all $\theta$ respecting $h$
and $\sigma \leq \theta$, it holds that $P\theta \vDash^{h\theta} \phi\theta$,
then $P \vDash^h \boxm{\sigma}\phi$ holds.
\end{lem}
\begin{lem}\label{lemma:respects}
If $\sigma \cdot \theta$ respects $h$,
then $\theta$ respects $h\sigma$.
\end{lem}
\begin{lem}[monotonicity]\label{lemma:mono-trans}
If $P \lts{\pi} Q$ then $P\theta \lts{\pi\theta} Q\theta$,
for all $\theta$ such that if $x \in \bn{\pi}$ and $y\theta = x$ then $x = y$.
\end{lem}
\begin{proof}
The proof follows directly by induction on the derivation of rules. We consider two cases only.
Consider the base case for input transitions.
\[
\infer{
a(x).P \lts{a(x)} P
}{}
\]
Clearly, if $x\theta = x$, then we have $a\theta(x).P\theta = (a(x).P)\mathclose{\theta}$ and the following labelled transition is enabled, as required.
\[
\infer{
a\theta(x).P\theta \lts{a\theta(x)} P\theta
}{}
\]
Consider the inductive case for the following rule.
\[
\infer{
\nu x.P \lts{\co{a}(x)} Q
}{
P \lts{\co{a}x} Q
}
\]
By the induction hypothesis, for all $\theta$, we have
$P\theta \lts{\co{a\theta}x\theta} Q\theta$.
Hence if, in addition, $x\theta = x$, we have
$P\theta \lts{\co{a\theta}x} Q\theta$ and $\nu x.P\theta = (\nu x.P)\theta$ and so the following labelled transition is enabled, as required.
\[
\infer{
\nu x.P\theta \lts{\co{a\theta}(x)} Q\theta
}{
P\theta \lts{\co{a\theta}x} Q\theta
} \qedhere
\]
\end{proof}

We comment on the generality of the results in this work.
Remarkably, monotonicity of the labelled transitions (Lemma~\ref{lemma:mono-trans}) is the only property we require of the process model in order to prove completeness, other than image finiteness (which, as discussed previously, could even be lifted in an extended logic).
Thus it would be possible to make the results of this paper more abstract, by ranging over any process model 
where the labelled transition system satisfies image finiteness (Lemma~\ref{lemma:finite}), monotonicity of the labelled transition system (Lemma~\ref{lemma:mono-trans}),
and, furthermore, has the same labels as for the late transition system of the $\pi$-calculus (i.e., of the form $\tau$, $\co{x}(z)$, $\co{x}z$ or $x(z)$, where $x$ and $z$ are variables).
When viewed in terms of Kripke semantics in Appendix~\ref{sec:kripke}, monotonicity is essentially the first compatibility condition of Plotkin and Sterling~\cite{Plotkin1986}; which is the ``zig-zag'' between the modal accessibility relation (here defined by the labelled transition along with its history) and the intuitionistic information partial ordering (here defined by substitutions respecting a history), that is sufficient to guarantee intuitionistic hereditary.

We have chosen to stick a more concrete formulation of the $\pi$-calculus rather than being more abstract, for two reasons.
Firstly, we can clearly provide concrete examples in a single process language familiar to a large audience.
Secondly, results obtained using such a more abstract approach should be treated with caution,
since it does not immediately cover many richer process calculi, for which the definition of open bisimilarity must be modified.
For example, when extending our results to the $\pi$-calculus with mismatch~\cite{Horne2018}, we require that the definition of open bisimilarity is extended to allow for the retrospective creation of fresh private names in the past, when our supply of private names runs out, otherwise open bisimilarity is not a congruence.
Going further, for the applied $\pi$-calculus~\cite{Abadi2018} or $\psi$-calculi~\cite{Bengtson2011},
to define open bisimilarity the labels employed are of a more general form, so both the labels and the definition of open bisimilarity change
in order to conservatively extend open bisimilarity to these settings while retaining the property that open bisimilarity is a congruence~\cite{Horne2020}. 
Thus it is a deliberate choice that, in this paper, we do not provide results at the maximum level of abstraction or generality that we know how to provide;
instead, we seek to clearly map out the key novel ideas in a widely understood process language.

\subsection{Algorithm for distinguishing formulae}%
\label{sec:complete:algo}
The direct definition of distinguishability (Definition~\ref{def:non-bisim}) provides us with a tree of substitutions and actions forming a strategy showing that two processes are not open bisimilar.
The following proposition shows that \OM\ formulae are sufficient to describe such strategies.
For any strategy that distinguishes two processes, we can construct \textit{distinguishing} formula in \OM.
A distinguishing formula holds for one process but not for the other process.
In the proof of the following proposition, 
at each step we construct two distinguishing formulae,
one biased to the process on the left and another biased to the process on the right, since we cannot simply construct a formula biased to one process and negate it to obtain a formula biased to the other process, which is the standard trick used since the early days of classical Hennessy-Milner logics~\cite{HenMil85}.
We discussed in Section~\ref{sec:direct}, why the left biased formula cannot be simply
obtained by negating the right biased formula and vice versa; both must be constructed simultaneously and
may be unrelated by negation.
\begin{prop}\label{proposition:non-bisim}
If $P \not\sim Q$ then there exists $\phi_L$ such that $P \vDash \phi_L$ and $Q \not\vDash \phi_L$.
\end{prop}
\begin{proof}
Since $\not\sim$ is defined by a least fixed point over a family of relations $\not\sim_n$,
if $P \not\sim^h Q$, there exists $n$ such that $P \not\sim_n^h Q$, so we can proceed by induction on the depth of a distinguishing strategy.

In the base case, assume $P \not\sim_0^h Q$, hence by definition,
 for substitution $\sigma$ respecting $h$,
$P\sigma \lts{\pi\sigma} P'$, for $x \in \bn{\pi}$, $x$ is fresh for $P\sigma$, $Q\sigma$ and $h\sigma$, such that there is no $Q'$ such that $Q\sigma \lts{\pi\sigma} Q'$.
It is sufficient to consider only this base case without loss of generality, since the other case is symmetric ($Q$ leads and $P$ cannot follow).

We require the following property concerning substitutions enabling $\pi\theta$-transitions from $Q\theta$, exploiting the observation that necessarily each such $\theta$ must induce an additional equality that was not yet enabled by $\sigma$.
There exist finitely many pairs of variables $x_j$ and $y_j$ in $\fv{P} \cup \fv{Q} \cup \fv{\pi}$ such that $x_j\sigma$ and $y_j\sigma$ are distinct, and, for any $R$ and substitution $\theta$ respecting $h$, if $Q\theta \lts{\pi\theta} R$ there exists $j$ such that $x_j \theta = y_j \theta$.
To see why, assume for contradiction that there is some $\theta$ respecting $h$ such that $Q\theta \lts{\pi\theta} R$ but there is no $x$ and $y$ in $\fv{P} \cup \fv{Q} \cup \fv{\pi}$ such that $x\sigma$ and $y\sigma$ are distinct, and $x \theta = y \theta$. 
Stated otherwise, 
for all $x$ and $y$ in $\fv{P} \cup \fv{Q} \cup \fv{\pi}$ if $x \theta = y \theta$ then $x\sigma = y\sigma$,
which is precisely the definition of a function, i.e.,\ there is a substitution, say $\theta'$, defined on the range of $\theta$
such that $\theta'$ maps $z \theta$ to $z\sigma$.
In that case, 
$\theta \cdot \theta' = \sigma$ on $\fv{P} \cup \fv{Q} \cup \fv{\pi}$; 
and hence, by Lemma~\ref{lemma:mono-trans}, 
$Q\theta \theta' \lts{\pi\theta \theta' } R\theta'$ contradicting the initial assumption for the base case that no transition $Q\sigma \lts{\pi\sigma} Q'$ exists for any $Q'$.

In this case, there are two distinguishing formulae $\boxm{\sigma}\diam{\pi}\ttt$ and $\boxm{\pi}\bigvee_j \left( x_j = y_j \right)$ biased to $P$ and $Q$ respectively.
There are four cases to check to confirm that these are distinguishing formulae.
\begin{description}[leftmargin=2.5mm]

\item[{Case} $P \vDash^h \boxm{\sigma}\diam{\pi}\ttt\,$]
Consider all $\theta$ respecting $h$ such that $\sigma \leq \theta$.
By definition there exists $\theta'$
such that $\sigma \cdot \theta' = \theta$,
so since $P\sigma \lts{\pi\sigma} P'$, by Lemma~\ref{lemma:mono-trans},
$P\theta \lts{\pi\theta} P'\theta'$.
Thereby, since $P'\theta' \vDash^{h'} \ttt$ holds,
$P\theta \vDash^{h\theta} \diam{\pi\theta}\ttt$.
Hence, by Lemma~\ref{lemma:box},
$P \vDash^h \boxm{\sigma}\diam{\pi}\ttt$.

\item[{Case} $Q \not\vDash^h \boxm{\sigma}\diam{\pi}\ttt\,$]
For contradiction, assume $Q \vDash^h \boxm{\sigma}\diam{\pi}\ttt$.
Since $\sigma$ respects $h$ and $\sigma \leq \sigma$, by Lemma~\ref{lemma:box}, $Q \vDash^h \boxm{\sigma}\diam{\pi}\ttt$ holds
only if $Q\sigma \vDash^{h\sigma} \diam{\pi\sigma}\ttt$ holds;
which holds only if there exists $Q'$ 
such that $Q\sigma \lts{\pi\sigma} Q'$, contradicting the assumption no such $Q'$ exists.
Thereby $Q \not\vDash^h \boxm{\sigma}\diam{\pi}\ttt$.

\item[{Case} $Q \vDash^h \boxm{\pi}\bigvee_j \left( x_j = y_j \right)\,$]
Consider substitutions $\theta$ respecting $h$ and $Q'$
such that $Q\theta \lts{\pi\theta} Q'$.
It must be the case that 
there exists $j$ 
such that $x_j \theta = y_j\theta$, 
thereby
$Q' \vDash^{h\theta}  x_j\theta = y_j\theta$ holds;
hence clearly $Q' \vDash^{h\theta} \bigvee_j \left( x_j = y_j \right)\mathclose{\theta}$ holds.
Hence $Q \vDash^h \boxm{\pi}\bigvee_j \left( x_j = y_j \right)$.

\item[{Case} $P \not\vDash^h \boxm{\pi}\bigvee_j \left( x_j = y_j \right)\,$]
Assume for contradiction $P \vDash^h \boxm{\pi}\bigvee_j \left( x_j = y_j \right)$.
This holds iff for all processes $S$ and substitutions $\theta$ respecting $h$,
$P\theta \lts{\pi\theta} S$ implies $S \vDash^{h'} \bigvee_j \left( x_j = y_j \right)\mathclose{\theta}$.
Since we know that $\sigma$ respects $h$ and $P\sigma \lts{\pi\sigma} P'$, for some $h''$,
we have $P' \vDash^{h''} \bigvee_j \left( x_j = y_j \right) \mathclose{\sigma}$.
This holds only if for some $j$, $P' \vDash^{h''}   x_j\sigma = y_j \sigma$;
hence, $x_j\sigma = y_j\sigma$ for some $j$, which contradicts the assumption
that $x_j\sigma$ and $y_j\sigma$ are distinct.
Thereby $P \not\vDash^h \boxm{\pi}\bigvee_j \left( x_j = y_j \right)$.
\end{description}

Now consider the inductive cases.
Given $P$, $Q$, if $P \not\sim_{n+1}^h Q$, up to symmetry of $\not\sim_{n+1}^h$, there are three cases to consider,
for some substitution $\sigma$ respecting $h$, where $\alpha$ is either $\tau$ or $\co{a}b$,
where $x$ is fresh for $P\sigma$, $Q\sigma$ and $h\sigma$:

\begin{itemize}
\item $P\sigma \lts{\alpha\sigma} P'$ and for all $Q_i$ such that $Q\sigma \lts{\alpha\sigma} Q_i$, $P' \not\sim^{h\sigma}_n Q_i$.
\item $P\sigma \lts{\co{a\sigma}(x)} P'$, and, for all $Q_i$ such that $Q\sigma \lts{\co{a\sigma}(x)} Q_i$, $P' \not\sim^{h\sigma \cdot x^o}_n Q_i$.
\item $P\sigma \lts{a\sigma(x)} P'$, and, for all $Q_i$ such that $Q\sigma \lts{a\sigma(x)} Q_i$, $P' \not\sim^{h\sigma \cdot x^i}_n Q_i$.
\end{itemize}
We consider the second case above involving bound output only, the other two cases are similar -- differing only in the accounting for respectful substitutions according to Def.~\ref{def:respects}.

For $P\sigma \lts{\co{a\sigma}(x)} P'$, by Lemma~\ref{lemma:finite}, there exist finitely many $Q_i$ 
such that $Q\sigma \lts{\co{a\sigma}(x)} Q_i$.
For each $i$, since $P' \not\sim^{h\sigma \cdot x^o}_n Q_i$, %$P' \not\sim_n^h Q_i$,
by the induction hypothesis, there exist $\phi^L_i$ and $\phi^R_i$ such that $P' \vDash^{h\sigma \cdot x^o} \phi^L_i \sigma$ and $Q_i \not\vDash^{h\sigma \cdot x^o} \phi^L_i \sigma$
and $P' \not\vDash^{h\sigma  \cdot x^o} \phi^R_i \sigma$ and $Q_i \vDash^{h\sigma \cdot x^o} \phi^R_i \sigma$.

We require the following property, referred to later using $\dagger$.
There are finitely many pairs of variables $x_j$ and $y_j$ selected from $\fv{P}\cup\fv{Q}\cup\left\{a\right\}$ such that $x_j\sigma$ and $y_j\sigma$ are distinct,
and, for any substitution $\theta$ respecting $h$ (note we can apply $\alpha$-conversion to ensure that $\theta$ also respects $h \cdot x^o$), such that $\sigma \leq \theta$, and for any $S$ such that, $Q\theta \lts{\co{a\theta}(x)} S$ then either:
for some $i$, we have $S \vDash^{h\theta \cdot x^o} \phi^R_i\theta$, 
or there exists some $j$ such that $x_j\theta = y_j\theta$.

To see why such pairs of variables $x_j$ and $y_j$ can be constructed, suppose, for contradiction, that they cannot be constructed in general.
Hence, there would exist substitution $\rho$ respecting $h\cdot x^o$, where $\sigma \leq \rho$, and process $S$ such that: $Q\rho \lts{\co{a\rho}(x)} S$, there is no $i$ such that 
$S \vDash^{h\rho \cdot x^o} \phi^R_i\rho$, and also there is no pair of variables $u$ and $v$ in $\fv{P}\cup\fv{Q}\cup\left\{a\right\}$ such that $u\sigma$ and $v\sigma$ are distinct and $u\rho = v\rho$. Hence $\rho \leq \sigma$; therefore, there exists $\rho'$ respecting $h\rho \cdot x^o$ such that $\rho\cdot\rho' = \sigma$ and hence, by Lemma~\ref{lemma:mono-trans}, $Q\sigma \lts{\co{a\sigma}(x)} S\rho'$, where $S\rho' = Q_i$ for some $i$.
Since, $\rho \leq \sigma$ and $\sigma \leq \rho$, we know $\rho'$ has an inverse, say $\sigma'$. Now since, by Lemma~\ref{lemma:mono-trans}, $Q\sigma\sigma' \lts{\co{a\sigma\sigma'}(x)} Q_i\sigma'$, we have $Q\rho \lts{\co{a\rho}(x)} Q_i\sigma'$; and, since $Q_i \vDash^{h\sigma \cdot x^o} \phi^R_i\sigma$, by Lemma~\ref{lemma:mono}, we have $Q_i\sigma' \vDash^{h\rho \cdot x^o} \phi^R_i\rho$, i.e.,
$S \vDash^{h\rho \cdot x^o} \phi^R_i\rho$,
 contradicting the assumption no such $i$ exists.

From the above, it is possible to construct distinguishing formulae $\boxm{\sigma}\diam{\co{a}(x)}\bigwedge_i \phi^L_i$ and
$\boxm{\sigma}\mathopen{\boxm{\co{a}(x)}}\left( \bigvee_i \phi^R_i \vee \bigvee_j \left(  x_j = y_j \right) \right)$.
There are four cases to consider to verify these are distinguishing formulae.
\begin{description}[leftmargin=2.5mm]

\item[{Case} $P \vDash^{h} \boxm{\sigma}\diam{\co{a}(x)}\bigwedge_i \phi^L_i$\,]
Consider all $\theta$ such that $\sigma \leq \theta$, $\theta$ respects $h$, and without loss of generality $x$ is fresh for $\theta$,\ i.e.,\ for $y \in \dom{\theta}$ and $x \not\in y\theta$.
By definition, there exists $\theta'$ such that $\sigma\cdot\theta' = \theta$.
Now since $\sigma\cdot\theta'$ respects $h$, by Lemma~\ref{lemma:respects}, $\theta'$ respects $h\sigma$
hence since $x \not\in \dom{\theta'}$ and $x \not\in \fv{h\sigma\theta'}$,
$\theta'$ respects $h\sigma \cdot x^o$.
Thereby 
since $\theta'$ respects $h\sigma \cdot x^o$ and
also $P' \vDash^{h\sigma\cdot x^o} \phi^L_i\sigma$ holds, by Lemma~\ref{lemma:mono},
it holds that $P'\theta' \vDash^{h\theta \cdot x^o} \phi^L_i \theta$.
The above holds for all $i$, hence it holds that 
$P'\theta' \vDash^{h\theta \cdot x^o} \bigwedge_i \phi^L_i\theta$.
Now, since $P\sigma \lts{\co{a\sigma}(x)} P'$, by Lemma~\ref{lemma:mono-trans}, since $x$ is fresh, $P\theta \lts{\co{a\theta}(x)} P'\theta'$ holds; 
and hence
$P\theta \vDash^{h\theta} \left(\diam{\co{a}(x)}\bigwedge_i \phi^L_i\right)\mathclose{\theta}$ holds.
Thereby, by Lemma~\ref{lemma:box}, 
$P \vDash^h \boxm{\sigma}\diam{\co{a}(x)}\bigwedge_i \phi^L_i$ holds.

\item[{Case} $Q \not\vDash^h \boxm{\sigma}\diam{\co{a}(x)}\bigwedge_i \phi^L_i$\,]
Assume for contradiction that
$Q \vDash^h \boxm{\sigma}\diam{\co{a}(x)}\bigwedge_i \phi^L_i$ holds.
Since $\sigma$ respects $h$ and $\sigma \leq \sigma$, by Lemma~\ref{lemma:box},
the above assumption holds 
only if $Q\sigma \vDash^{h\sigma} \left(\diam{\co{a}(x)}\bigwedge_i \phi^L_i\right)\mathclose{\sigma}$ holds.
Now $Q\sigma \vDash^{h\sigma} \diam{\co{a\sigma}(x)}\bigwedge_i \phi^L_i\mathclose{\sigma}$ holds 
only if there exists $Q'$ such that
$Q\sigma \lts{\co{a\sigma}(x)} Q'$ and 
$Q' \vDash^{h\sigma \cdot x^o} \bigwedge_i \phi^L_i \sigma$,
which holds only if
$Q' \vDash^{h\sigma \cdot x^o} \phi^L_i \sigma$ for all $i$.
Notice that $Q' = Q_k$ for some $k$,
and therefore $Q_k \vDash^{h\sigma \cdot x^o} \phi^L_k \sigma$;
but it was assumed that $Q_k \not\vDash^{h\sigma \cdot x^o} \phi^L_k \sigma$ leading to a contradiction.
Therefore $Q \not\vDash^h \boxm{\sigma}\diam{\co{a}(x)}\bigwedge_i \phi^L_i$.

\item[{Case} $Q \vDash^{h} \boxm{\sigma}\boxm{\co{a}(x)}\left(\bigvee_i \phi^R_i \vee \bigvee_j \left(  x_j=y_j \right) \right)$\,]
Fix $Q'$ and $\theta$ respecting $h$, 
such that $\sigma \leq \theta$, and $Q\theta \lts{\co{a\theta}(x)} Q'$.
Above, in $\dagger$, we established that, in this scenario, 
either: for some $\ell$, we have $Q' \vDash^{h\theta \cdot x^o} \phi^R_\ell\theta$, 
or there exists some $k$ such that $x_k\theta = y_k\theta$.
In the case where, for some $k$, $x_k\theta = y_k\theta$, 
we have $Q' \vDash^{h\theta \cdot x^o} {x_k\theta=y_k\theta}$ holds.
Hence in either case we have
$Q' \vDash^{h\theta \cdot x^o} \left(\bigvee_i \phi^R_i \vee \bigvee_j \left({x_j=y_j}\right)\right)\mathclose{\theta}$, by definition of disjunction.
Thereby, by definition,
$Q \vDash^h \boxm{\sigma}\boxm{\co{a}(x)}\left(\bigvee_i \phi^R_i \vee \bigvee_j \left({x_j=y_j}\right)\right)$ holds.

\item[Case $P \not\vDash^h \boxm{\sigma}\boxm{\co{a}(x)}\left(\bigvee_i \phi^R_i \vee \bigvee_j \left({x_j=y_j}\right)\right)$\,]
Let us assume for contradiction that $P \vDash^h \boxm{\sigma}\boxm{\co{a}(x)}\left(\bigvee_i \phi^R_i \vee \bigvee_j \left({x_j=y_j}\right)\right)$.
Since $\sigma$ respects $h$, $\sigma \leq \sigma$, and $P\sigma \lts{\co{a\sigma}(x)} P'$,
the previous assumption can hold only if
$P' \vDash^{h\sigma\cdot x^o} \left(\bigvee_i \phi^R_i \vee \bigvee_j \left({x_j=y_j}\right)\right)\mathclose{\sigma}$.
This holds only if, for some $i$, $P' \vDash^{h\sigma\cdot x^o} \phi^R_i \sigma$, or, for some $j$, $P' \vDash^{h\sigma\cdot x^o} \left({x_j\sigma=y_j\sigma}\right)$.
However, for all $i$, $P' \not\vDash^{h\sigma\cdot x^o} \phi^R_i \sigma$;
and also, for all $j$, we have $x_j\sigma$ and $y_j\sigma$ are distinct and $P' \not\vDash^{h\sigma\cdot x^o} \left({x_j\sigma=y_j\sigma}\right)$, leading to a contradiction in either case.
Thereby $P \not\vDash^h \boxm{\sigma}\boxm{\co{a}(x)}\left(\bigvee_i \phi^R_i \vee \bigvee_j \left({x_j=y_j}\right)\right)$.
\end{description}

By induction we have established that, for any history $h$, processes $P$ and $Q$, and any $n$,
if $P \not\sim^h_n Q$ then we can construct $\phi_L$ such that $P \vDash^h \phi_L$ and $Q \not\vDash^h \phi_L$; and also 
we can construct $\phi_R$ such that $Q \vDash^h \phi_R$ and $P \not\vDash^h \phi_R$.
The result then follows by observing that, since $\not\sim$ is the least relation containing all $\not\sim_n$ whenever $P \not\sim Q$;
there exists $n$ such that $P \not\sim^{x_1^i\cdot\hdots\cdot x^i_n}_n Q$ and, where $\fv{P}\cup\fv{Q} \subseteq \left\{ x^i_1, \hdots, x^i_n \right\}$;
for which, there is $\phi^L$ such that $P \vDash^{x_1^i\cdot\hdots\cdot x^i_n} \phi_L$ and $Q \not\vDash^{x_1^i\cdot\hdots\cdot x^i_n} \phi_L$; and also $\phi^R$ such that $Q \vDash^{x_1^i\cdot\hdots\cdot x^i_n} \phi_R$ and $P \not\vDash^{x_1^i\cdot\hdots\cdot x^i_n} \phi_R$.
Hence, by Definition~\ref{def:modal}, 
indeed $P \vDash \phi_L$, $Q \not\vDash \phi_L$, $Q \vDash \phi_R$ and $P \not\vDash \phi_R$ as required.
\end{proof}

\subsection{The proof of completeness}%
\label{sec:complete:proof}

Combining Proposition~\ref{proposition:non-bisim} with Lemma~\ref{lemma:not} yields immediately the completeness of \OM\ with respect to open bisimilarity.
Completeness (Theorem~\ref{theorem:complete}) establishes that the set of all pairs of processes that have the same set of distinguishing formulae 
is an open bisimulation. The proof can now be stated as follows.

\textit{Proof of Theorem~\ref{theorem:complete}:}
Assume that for finite processes $P$ and $Q$, for all formulae $\phi$, $P \vDash \phi$ iff $Q \vDash \phi$.
Now for contradiction suppose that $P \sim Q$ does not hold.
By Lemma~\ref{lemma:not}, $P \not\sim Q$ must hold.
Hence by Proposition~\ref{proposition:non-bisim}
there exists $\phi_L$ such that
$P \vDash \phi_L$ but $Q \not\vDash \phi_L$, but we assumed at the beginning that $P \vDash \phi_L$ holds iff $Q \vDash \phi_L$ holds, leading to a contradiction.
Thereby $P \sim Q$.
\hfill\qed

\subsection{Example runs of distinguishing formula algorithm}\label{section:egs}
We provide further examples of processes that are not open bisimilar
that illustrate subtle aspects of the algorithm.
In particular, these examples illustrate various scenarios where postconditions are required.

\subsubsection{Multiple postconditions and postconditions in an inductive step}\label{eg:multiple}\label{eg:inductive}

The following example leads to multiple postconditions.
Consider the following distinguishable processes.
\[
\match{x=y}\tau + [w=z]\tau \qquad\not\sim\qquad \tau
\]
Observe that clearly $\tau \lts{\tau} 0$ but $\left(\match{x=y}\tau + [w=z]\tau\right)\mathclose{\theta} \lts{\tau}$
only if $x\theta = y\theta$ or $w \theta = z\theta$.
Thus, 
	$\match{x=y}\tau + [w=z]\tau \vDash \mathopen{\boxm{\tau}}\left((x=y)\vee(w=z)\right)$
is a distinguishing formula biased to the left process,
while $\tau \vDash \diam{\tau}\ttt$ is biased to the right.\\[-1.8ex]

We consider now an example where postconditions are required in the inductive case of the distinguishing formulae algorithm.
However, firstly observe that $\co{a}a + \co{b}b \not\sim \co{a}a$ are distinguished since
$\co{a}a + \co{b}b \lts{\co{b}b} 0$, but process $\co{a}a$ can only make a $\co{b}b$ transition under a substitution such that $a = b$.
Hence we have the distinguishing formulae
$\co{a}a + \co{b}b \vDash \diam{\co{b}b} \ttt$ and $\co{a}a \vDash \boxm{\co{b}b}(a=b)$.
Now consider the following.
\[
 P\triangleq \mathopen{\tau.}\left(\co{a}a + \co{b}b\right) + \match{x=y}\tau.\co{a}a
\qquad \not\sim \qquad
\mathopen{\tau.}\left(\co{a}a + \co{b}b\right) + \tau.\co{a}a \triangleq Q 
\]
To distinguish these processes, $Q \lts{\tau}\co{a}a$ leads, a move which can only be matched by
$P \lts{\tau} \co{a}a + \co{b}b$. 

To construct formulae distinguishing $P$ from $Q$ we use the following ingredients:
the distinguishing formulae constructed for the sub-problem $\co{a}a + \co{b}b \not\sim \co{a}a$;
and the observation that, for substitutions $\theta$ such that $x\theta = y\theta$, there is an additional $\tau$-transitions enabled: $P{\theta} \lts{\tau} \co{a}a$.
These observations lead us to the following distinguishing formula biased to the process $P$ in the left above, consisting of a box $\tau$ 
followed by a disjunction comprised of the distinguishing formula for $\co{a}a + \co{b}b \not\sim \co{a}a$ biased to the process $\co{a}a + \co{b}b$ on the left,
and the postcondition $x = y$, which must hold after the additional $\tau$ transition is enabled.
\[
\mathopen{\tau.}\left(\co{a}a + \co{b}b\right) +
\match{x=y}\tau.\co{a}a
\vDash
\mathopen{\boxm{\tau}}\left( \diam{\co{b}b}\ttt \vee x=y \right)
\]
The distinguishing formula biased to the process $Q$ on the right above is ``diamond $\tau$'' 
followed by the distinguishing formula for $\co{a}a + \co{b}b \not\sim \co{a}a$ biased to the process $\co{a}a$, as follows.
\[
\mathopen{\tau.}\left(\co{a}a + \co{b}b\right) +
\tau.\co{a}a
\vDash
\diam{\tau}\boxm{\co{b}b}(a=b)
\]

\subsubsection{Formulae generated by substitutions applied to labels}\label{eg:subslabel}
In some cases, substitutions applied to labels play a role when generating distinguishing formulae.
For a minimal example consider the following distinguishable processes: $
\co{a}a  \not\sim  \co{a}b
$.
A distinguishing strategy is where process $\co{a}b$ makes a $\co{a}b$ transition, which cannot be matched by $\co{a}a$. However, we do have transition $\left(\co{a}a\right)\mathclose{\sigma} \lts{\left(\co{a}b\right)\mathclose{\sigma}} 0$ for any substitution such that $a\sigma = b\sigma$, leading to distinguishing formula $\boxm{\co{a}b}(a=b)$ biased to $\co{a}a$.
Notice substitution $\sigma$ is applied to both the process and the label.

For a trickier example consider the following processes.
\[
\nu b.\co{a}b.a(x).\match{x=b}\co{x}x
\not\sim
\nu b.\co{a}b.a(x).\co{x}x
\]
After two actions, the above problem reduces to base case
$\match{x=b}\co{x}x \not\sim^{a^i\cdot b^o\cdot x^i}\co{x}x$,
where $\co{x}x$ can perform a $\co{x}x$ action, but $\match{x=b}\co{x}x$ cannot.
However, $\left(\match{x=b}\co{x}x\right)\sub{x}{b} \lts{\co{x}x\sub{x}{b}} 0$ does hold,
and furthermore $\sub{x}{b}$ respects $a^i\cdot b^o\cdot x^i$.
From these observations we can construct a distinguishing formula biased to the left as follows.
\[
\nu b.\co{a}b.a(x).\match{x=b}\co{x}x
\vDash
\boxm{a(b)}\boxm{a(x)}\boxm{\co{x}x}(x=b)
\]

\subsubsection{
Alternative forms for distinguishing formulae
}\label{eg:human}
Note our algorithm copes with sub-optimal distinguishing strategies.
To understand this, consider the distinguishing strategy for the following processes that are clearly not open bisimilar.
\[
\match{x = y}\tau  
\qquad
\not\sim
\qquad
\tau.\match{x = y}\tau  
\]
There is an obvious optimal distinguishing strategy: $\tau.\match{x = y}\tau \lts{\tau} \match{x = y}\tau$, which cannot be matched by $\match{x = y}\tau$.
By appealing to the base case of the distinguishing formulae algorithm, we obtain two distinguishing formulae $\boxm{\tau}\left(x = y\right)$ and $\diam{\tau}\ttt$ biased each respective process.

There are however, sub-optimal, distinguishing strategies. Under substitution $\sub{x}{y}$, the process on the left has transition $\left(\match{x = y}\tau\right)\sub{x}{y} \lts{\tau} 0$,
which can be matched, under the same substitution, by $\left(\tau.\match{x = y}\tau\right)\sub{x}{y} \lts{\tau} \match{y = y}\tau$. Now $0$ and $\match{y = y}\tau$ are distinguished, since $\match{y = y}\tau \lts{\tau} 0$ whereas $0$ is deadlocked. By applying the algorithm in Proposition~\ref{proposition:non-bisim},
we obtain the formula $x = y \yields \boxm{\tau}\diam{\tau}\ttt$ biased to the process on the right, which is indeed distinguishing.

As a further example of alternative distinguishing formulae, consider the following processes.
\[
\boxm{x=y}\tau.\tau + \tau
\qquad
 \not\sim 
\qquad
\tau.\tau + \tau
\]
The following is a distinguishing formula biased to the left process: $\boxm{\tau}\boxm{\tau}(x=y)$.
However, this is different from the left-biased formula $\mathopen{\boxm{\tau}}\left( \boxm{\tau}\fff \vee (x=y) \right)$ generated by the algorithm.
Thus, there exist alternative distinguishing formulae \ldots and alternative algorithms.
In particular, the above two examples highlight the open question of whether restricting ourselves to minimal substitutions in the distinguishing strategy allows us to simplify slightly the formulae in the inductive case of the distinguishing formula algorithm, thereby avoiding generating formulae such as 
$x = y \yields \boxm{\tau}\diam{\tau}\ttt$ featuring a prefix $x = y \yields$ before a box modality. 

\subsubsection{A more elaborate example}\label{eg:elaborate}
 This example forces the use of postconditions regardless of whether we construct a distinguishing formula biased to the process on the left or on the right.
Consider the following processes.
\[
P \triangleq \tau  + \mathopen{\tau.}\left( \tau.\tau + \tau \right) + \mathopen{\tau.\match{x = y}}\left( \tau.\match{u = v}\tau + \tau.\tau + \tau \right)
\qquad
Q \triangleq P + \mathopen{\tau.\match{x = y}}\left( \tau.\tau + \tau \right)
\]
The processes above are distinguished by the following strategy.
Firstly, the process $Q$ moves, as follows; for which there are three moves $P$ can perform.

\qquad
\xymatrix@R=8pt{
Q\ar^{\tau}[d]
&&
&
P \ar^{\tau}[d]\ar@/_/_{\tau}[dl]\ar@/^/^{\tau}[dr]
\\
\match{x = y}\left( \tau.\tau + \tau \right)
&&
0
&
\tau.\tau + \tau
&
\match{x = y}\left( \tau.\match{u = v}\tau + \tau.\tau + \tau \right)
}

This leads to three sub-problems, for which we know already the distinguishing strategies and formulae.
Note, to distinguish 
$\match{x = y}\left( \tau.\tau + \tau \right)$ from $\match{x = y}\left( \tau.\match{u = v}\tau + \tau.\tau + \tau \right)$,
there is a switch in the process that leads.

From the above strategy, we can construct the following distinguishing formulae.
\[
Q \vDash \diam{\tau}\left( \left( x = y \yields \diam{\tau}\ttt \right) \wedge \boxm{\tau}\left( x = y \right) \wedge \boxm{\tau}\left( \diam{\tau}\ttt \vee \boxm{\tau}\fff \right) \right)
\]
\[
P \vDash \boxm{\tau}\left( \diam{\tau}\ttt \vee \boxm{\tau}\fff \vee \left(x = y \yields \diam{\tau}\left( \left( u = v \yields \diam{\tau}\ttt \right) \wedge \boxm{\tau}\left( u = v \right)   \right)  \right) \right)
\]
Notice this example nests a classic example, explained previously, inside itself.
The absence of the law of excluded middle is essential for the existence of distinguishing formulae in this example.

\section{Situating $\mathcal{OM}$ with Respect to Other Modal Logics Characterising Bisimilarities} \label{sec:related}

Open bisimilarity is not the only bisimilarity congruence.
We consider here the relationship between the intuitionistic modal logic for open bisimilarity presented in this work and other modal logics.
In doing so, we clarify why we introduce \OM\ rather than taking an intuitionistic variant of an established modal logic.
We check that \OM\ has a classical counterpart characterising late bisimilarity.
Also, we note open bisimilarity is not the only notion of bisimilarity that is a congruence relation.
We provide a sharp picture explaining where open bisimilarity sits in relation to other notions of bisimilarity;
notably the bisimilarity congruences \textit{open barbed bisimilarity}~\cite{Sangiorgi2001} and a newly introduced late variant of open barbed bisimilarity which we call \textit{intermediate bisimilarity}, both of which can also be characterised by intuitionistic modal logics.

\subsection{Why a new modal logic \OM, rather than an intuitionistic variant of $\LM$?} 
\label{sec:related:nabla}

A classical logic characterising \textit{late bisimilarity}, called $\LM$ for ``($\mathcal{L}$) late modality with ($\mathcal{M}$) match,'' was provided by Milner, Parrow, and Walker~\cite{Milner93}.
$\LM$ differs from $\OM$ in two significant ways.
Firstly, $\LM$\ is classical: a classical semantics is induced due to the fact that all grounded inputs are considered immediately after an input action, where variables appearing free represent distinct ground names, hence an input is either equal to another ground message or it is not.
Secondly, the late input box modality is defined differently, involving an existential quantification over substitutions.
Moving to an intuitionistic variant of $\LM$, this gives rise to the following variant of the box input modality. 
\[
  % prevent line break before \phi at the end
  \text{
 $P \vDash \boxm{a(x)}^L\phi$
 iff $\forall \sigma \respecting h, \forall Q, P\sigma \lts{a\sigma(x)} Q ==> \exists x, \mbox{such that } Q \vDash^{h\sigma} \phi$.
  }
\]
In the semantics of \OM, we deliberately use a universally quantified box input modality, recalled bellow; rather than existentially quantified box input modality used in $\LM$ above.
\begin{quote}
 $P \vDash \boxm{a(x)}\phi$
 iff $\forall \sigma \respecting h, \forall Q, P\sigma \lts{a\sigma(x)} Q ==> Q \vDash^{h\sigma\cdot x^i} \phi$.
\end{quote}
Recall from Sec.~\ref{sec:abella}, in the box input modality of \OM\ immediately above, the $x^i$ appended to the history has the effect of $\forall x$ appearing immediately after the implication (made explicit in Fig.~\ref{fig:BisimSatDefs}).

Hence, due to the differences in quantification for the box input modality, \OM\ is not quite an intuitionistic variant of $\LM$.
The carefully selected box input modality in \OM\ is necessary for our construction of distinguishing formulae in the completeness proof for the characterisation of open bisimilarity using \OM.
To understand why, consider the following processes that are not open bisimilar.
\[
a(x).\tau + a(x) + a(x).[x=a]\tau
\quad
\not\sim
\quad
a(x).\tau + a(x)
\]
For the above processes, our algorithm for distinguishing formulae, Proposition~\ref{proposition:non-bisim},
correctly generates the following \OM\ formula biased to the right:
\[
a(x).\tau + a(x) + a(x).[x\!=\!a]\tau  \!\not\vDash\!
\boxm{a(x)}\left(\diam{\!\tau\!}\ttt \!\vee\! \boxm{\tau}\fff\right)
~~\,
\mbox{and}
~~\,
a(x).\tau + a(x)  \vDash  \boxm{a(x)}\left(\diam{\tau}\ttt \!\vee\! \boxm{\tau}\fff\right)
\]

If we were to use an intuitionistic variant of the input box modality of $\LM$, as suggested in related work~\cite{TiuMil10},
both processes satisfy the above formulae modified with a late box input modality $\mathopen{\boxm{a(x)}^{L}}\left(\diam{\tau}\ttt \vee \boxm{\tau}\fff\right)$.
To see why
$a(x).\tau + a(x) + a(x).[x=a]\tau \vDash \mathopen{\boxm{a(x)}^{L}}\left(\diam{\tau}\ttt \vee \boxm{\tau}\fff\right)$ holds,
observe that for the transition $a(x).\tau + a(x) + a(x).[x=a]\tau \lts{a(x)} [x=a]\tau$ there exists $\sub{x}{a}$ such that
$\left([x=a]\tau\right)\sub{x}{a} \vDash \diam{\tau}\ttt \vee \boxm{\tau}\fff$.

Also, note the formula $\mathopen{\boxm{a(x)}^{L}}\left(\left(x=a \wedge \boxm{\tau}\fff\right) \vee \diam{\tau}\mathopen{\neg}\left(x=a \right)\right)$ fails in intuitionistic $\LM$ for both processes, despite being distinguishing for these processes in classical $\LM$.
Although, if we use triple negation the formula becomes distinguishing in intuitionistic $\LM$, as follows:
\begin{equation}
a(x).\tau + a(x)  \vDash \mathopen{\boxm{a(x)}^{L}}\left(\left(x=a \wedge \boxm{\tau}\fff\right) \vee \diam{\tau}\mathopen{\neg\neg\neg}\left(x=a \right)\right)
\label{eq:triple-neg}
\end{equation}

\subsubsection{Example where box input is necessary.}
Notice, for the above example processes, formula $\diam{a(x)}\left(x = a \yields \diam{\tau}\ttt \wedge \boxm{\tau}\left(x = a\right)\right)$ is a distinguishing formula biased to the left, in both intuitionistic $\LM$ and \OM, since the diamond input modalities are the same in both intuitionistic modal logics.
For a more sophisticated example where box input modalities are necessary, regardless of the bias, consider the follow processes that are not open bisimilar.
\[
P \triangleq a(x) + \mathopen{a(x).}\left( a(y) + a(y).\tau \right) + a(x).\match{x = v}\left( a(y) + a(y).\tau + a(y).\match{y = w}\tau \right)
\]
\[
Q \triangleq P + a(x).\match{x = v}\left( a(y) + a(y).\tau \right)
\]
Consider the distinguishing strategy.
The first move must be made by $Q$, which can be matched by $P$ in three ways.

~\xymatrix@R=7pt{
 Q \ar^{a(x)}[d] 
&
& P\ar^{a(x)}[d]\ar@/_/_{a(x)}[dl]\ar@/^/^{a(x)}[dr]    &
\\
 \!\!\!\!\!\!\!\match{x = v}\left( a(y) + a(y).\tau \right)
&
\!\!\!\!\!\!\!\!\!0
&
 \!\!\!\!\!\!\!a(y) + a(y).\tau
&
\!\!\!\!\!\!\!\!\!\!\!\match{x = v}\left( a(y) + a(y).\tau + a(y).\match{y = w}\tau \right)
}

Each of the processes reachable by a $a(x)$-transition from $P$ are not open bisimilar to the process reachable from $Q$ indicated above.
The interesting case is the third process reached from $P$ above. After applying substitution $\sub{x}{v}$, the process on the right leads the distinguishing strategy.

\xymatrix@R=7pt{
&\match{v = v}\left( a(y) + a(y).\tau \right)\ar@/_/^{a(y)}[dl]\ar@/^/_{a(y)}[dr]
&&
&
\match{v = v}\left( a(y) + a(y).\tau + a(y).\match{y = w}\tau \right)\ar^{a(y)}[d]
\\
0& & \tau
&&
\match{y = w}\tau
}

The necessity of box input modalities is due to the switch from $Q$ leading initially to the other process leading for the second input in the distinguishing strategy.
From the above distinguishing strategy the following formula biased to $P$ can be constructed.
\[
P \vDash \boxm{a(x)}\left( \boxm{a(y)}\fff \vee \diam{a(y)}\ttt \vee \left( x = v \yields \diam{a(y)}\left( y = w \yields \diam{\tau}\ttt \wedge \boxm{\tau}\left( y = w \right) \right) \right) \right)
\]
For a formula biased to $Q$ we obtain the following.
\[
Q \vDash \diam{a(x)}\left( x = v \yields \diam{a(y)}\ttt \wedge \boxm{a(y)}\left( x = v\right) \wedge \boxm{a(y)}\left( \diam{\tau}\ttt \vee \boxm{\tau}\fff \right) \right)
\]
Neither of the above formulae would be distinguishing if, instead of the open box modalities of \OM, the late box modalities of $\LM$ were employed.

\subsubsection{Discussion on intuitionistic $\LM$.}
We have formalised the intuitionistic variant of $\LM$ in Abella. The language of formulae for $\LM$ replaces the ``basic'' box input modality of $\OM$ with
the following ``late'' box input modality:
\begin{lstlisting}
Type boxInL      n -> (n -> o') -> o'.
\end{lstlisting} 
The clauses for the satisfaction relation (encoded as the predicate \lstinline|satLM|) are those for $\OM$ (Figure~\ref{fig:BisimSatDefs}) without the ``basic'' box operator, 
but with the following clause for \lstinline|satLM| :
\begin{lstlisting}
satLM P (boxInL X A) := forall Q, {oneb P (dn X) Q} ->  exists z, sat (Q z) (A z). 
\end{lstlisting}
The example involving triple negation above (\ref{eq:triple-neg}) has been verified using this formalisation of intuitionistic $\LM.$

Related work~\cite{TiuMil10} suggested that intuitionistic $\LM$ characterises open bisimilarity. 
Unfortunately, the completeness proof in that work is flawed since they appeal to classical principles that are not valid in the intuitionistic setting.
This oversight is rectified in the current paper, by a more direct construction in the completeness proof and by the careful choice of input modalities in $\mathcal{OM}$, explained in this section.
Note however the example above involving triple negation, suggests the problem of whether intuitionistic $\LM$ characterises open bisimilarity remains an open problem.
To offer an intuition for triple negation: it can be regarded as an explicit test that variables are ``not equal yet,'' in contrast to single negation indicating that variables are never going to be equal.

\subsection{What about the classical counterpart to \OM?}
A criteria an intuitionistic modal logic is expected to satisfy is that, when the law of excluded middle is induced, we obtain a meaningful classical logic~\cite{Simpson94}.
Fortunately, this criteria holds for \OM\ -- the classical counterpart to \OM\ characterises late bisimilarity.
For convenience we, recall a definition of late bisimilarity.
\begin{defi}[late bisimilarity]
A late bisimulation $\mathcal{R}$ is a symmetric relation, such that, whenever $P \mathrel{\mathcal{R}} Q$:
\begin{itemize}
\item If $P \lts{\alpha} P'$ then there exists $Q'$ such that $Q \lts{\alpha} Q'$ and $P'  \mathrel{\mathcal{R}} Q'$.
\item If $P \lts{\co{a}(x)} P'$ then there exists $Q'$ such that $Q \lts{\co{a}(x)} Q'$ and $P'  \mathrel{\mathcal{R}} Q'$.
\item If $P \lts{a(x)} P'$ then there exists $Q'$ such that $Q \lts{a(x)} Q'$ and, for all $x$, $P'  \mathrel{\mathcal{R}} Q'$.
\end{itemize}
Late bisimilarity $\lsim$ is the greatest late bisimulation.
\end{defi}

\begin{figure}[h!]
\[
\begin{array}{c}
\!\!\!\!\!\!\!\!\!\!\!\!\!\!\!\!\!\!\!\!\!\!\!\!\!\!\!\!\!\!\!\!\!\!\!\!\!
\\
\begin{array}{lcl}
P \Late \ttt &\mbox{and}&\qquad P \Late x=x \qquad \mbox{always hold.}
\\
P \Late \phi_1 \yields \phi_1 &\mbox{iff}&
 P \Late \phi_1 ~ ==> ~ P \Late \phi_2.
\\
P \Late \diam{\alpha}\phi &\mbox{iff}&
  \exists\,Q,~ P \lts{\alpha} Q ~\mbox{and}~ Q \Late \phi.
\\
P \Late \diam{{a}(z)}\phi &\mbox{iff}&
  \exists\,Q,~ P \lts{{a}(z)} Q ~\mbox{and,}~ 
  \forall y,\, Q\sub{z}{y} \Late \phi\sub{z}{y}.
\\
 P \Late \boxm{{a}(z)}\phi &\mbox{iff}&
  \forall Q, 
    P \lts{a(z)} Q  ==> \forall z, Q \Late \phi.
\end{array}
\end{array}
\]
\caption{Semantics of ``classical \OM'', where $\alpha$ is $\tau$, $\co{a}b$ or $\co{a}(z)$.}
\label{figure:classicalOM}
\end{figure}

A direct semantics of classical \OM, in the style of Milner, Parrow and Walker~\cite{Milner93}, is presented in Fig.~\ref{figure:classicalOM}.
Observe histories are not employed in the classical semantics since inputs are instantiated eagerly, immediately after performing an input transition (see the clauses for the input labelled transitions).
Also, missing operators (conjunction, disjunction, and $\boxm{\alpha}\phi$) are derivable using classical negation; whereas in an intuitionistic modal logic they have independent interpretations.
Classical \OM\ characterises late bisimilarity.
\begin{cor}[characterisation]\label{theorem:classical}
$P \lsim Q$ if and only if, for all \OM\ formulae $\phi$, we have $P \Late \phi$ iff $Q \Late \phi$, according to the classical semantics for \OM\ in Fig.~\ref{figure:classicalOM}.
\end{cor}
\begin{proof}
Observe that the definition of $\diam{a(x)}\phi$ in Figure~\ref{figure:classicalOM} coincides with the late modality $\diam{a(x)}^L \phi$ in 
$\LM$. 
Also observe that, classically, $\neg\boxm{a(x)}\neg\phi$ is the ``basic'' diamond modality of Milner, Parrow and Walker~\cite{Milner93}; hence classical $\mathcal{OM}$ is classical $\LM$ extended with ``basic'' modalities.
That original paper on modal logics for the $\pi$-calculus establishes that, classical $\LM$ characterises late bisimilarity, and also
$\LM$ extend with basic modalities has the same expressive power at $\LM$.
\end{proof}

Historically, Milner, Parrow and Walker emphasised late equivalence (the greatest congruence contained in late bisimilarity) rather than late bisimilarity in the original paper on the $\pi$-calculus\cite{Milner92}. This is because late equivalence is closed under input prefixes.
 Late equivalence can be defined by restricting to late bisimulations closed  under substitutions;
and its characterisic modal logic can be defined in a simlar way,
 as follows.
\begin{defi}
$P$ is late equivalent to $Q$, written $P \mathrel{{\sim}_L} Q$, whenever there exists a late bisimulation $\mathcal{R}$ such that for all $\sigma$, $P\sigma \mathrel{\mathcal{R}} Q\sigma$.
Define $P \mathrel{{\vDash}_L} \phi$ whenever for all $\sigma$, $P\sigma \Late \phi\sigma$.
\end{defi}
Quantifying over all substitutions, combined with the distinct name assumption, means that we check late bisimilarity with respect to all combinations of equalities and inequalities between free variables.
As such, late equivalence is not a bisimilarity; but is a late bisimulation.
Using the above, we obtain a characteristic logic for late equivalence, using \OM\ formulae.
\begin{cor}\label{cor:late}
$P \mathrel{{\sim}_L} Q$ if and only if, for all $\phi$, $P \mathrel{{\vDash}_L} \phi$ iff $Q \mathrel{{\vDash}_L} \phi$.  
\end{cor}
\begin{proof}
This follows immediately from Corollary~\ref{theorem:classical} and the following facts. For $a$ fresh for $P$, $Q$ and $\phi$, $a(x_1)\hdots a(x_n).P \lsim a(x_1)\hdots a(x_n).Q$ if and only if $P \mathrel{{\sim}_L} Q$, and $P \mathrel{{\vDash}_L} \phi$ if and only if $a(x_1)\hdots a(x_n).P \Late \diam{a(x_1)}\hdots \diam{a(x_n)}\phi$ where $\fv{P} \cup \fv{Q} \cup \fv{\phi} \subseteq \left\{ x_1, \hdots, x_n \right\}$. 
\end{proof}

 As for open bisimilarity, $\match{x=y}\tau$ and $0$ are not late equivalent.
 This is because $\left(\match{x=y}\tau\right)\sub{y}{x}$ and $0\sub{y}{x}$ are clearly not late bisimilar.
Two distinguishing formulae in this logic are defined as follows:
$P \mathrel{{\vDash}_L} x = y \yields \diam{\tau}\ttt$
and
$Q \mathrel{{\vDash}_L} \boxm{\tau}\fff$.

The point is, if we take $\OM$\ and induce the law of excluded middle, we obtain a logic, defined by ${\vDash}_L$, characterising late equivalence.
If we then, in addition, enforce the distinct name assumption, we obtain a logic, defined by $\Late$, characterising late bisimilarity.

\subsection{A sharpened picture of the spectrum of bisimilarity congruences}
\label{sec:related:mpw}
We emphasise here that open bisimilarity is not the only bisimilarity congruence.
A notable, strictly coarser, bisimilarity congruence for the $\pi$-calculus is \textit{open barbed bisimilarity}~\cite{Sangiorgi2001}.
Notions of open barbed bisimilarity are, by definition, the greatest bisimilarity congruences. We give the strong formulation of open barbed bisimilarity here, consistent with the rest of the paper (of course, the weak formulation of open barbed bisimilarity is coarser).
\begin{defi}
Process $P$ has a barb $x$, written $\barb{P}{x}$, whenever $P \lts{x(z)} P'$ or $P \lts{\co{x}y} P'$ or $P \lts{\co{x}(z)} P'$.
An \textit{open barbed bisimulation} $\mathcal{R}$ is a symmetric relation such that, whenever $P \mathrel{\mathcal{R}} Q$ we have:
\begin{itemize}
\item If $P \lts{\tau} P'$ then there exists $Q'$ such that $Q \lts{\tau} Q'$ and $P'  \mathrel{\mathcal{R}} Q'$.
\item If $\barb{P}{x}$ then $\barb{Q}{x}$.
\item For contexts $\context{\ \cdot\ }$, we have $\context{P} \mathrel{\mathcal{R}} \context{Q}$.
\end{itemize}
Open barbed bisimilarity $\qsim$ is the greatest open barbed bisimulation.
\end{defi}
Unlike open bisimilarity, open barbed bisimilarity is incomparable with late bisimilarity.
A key example that holds for open barbed bisimilarity, but not for late bisimilarity is the following.
\[
\mathopen{\nu k.\co{a}k.}\left( a(x).P + a(x) \right) \quad  \qsim \quad \mathopen{\nu k.\co{a}k.}\left( a(x).\match{x = k}P  +  a(x).P + a(x) \right)
\]

There is however a (minimal) refinement of open barbed bisimilarity forbidding the above property, defined as follows.
\begin{defi}\label{definition:isim}
An intermediate bisimulation $\mathcal{R}$ is a symmetric relation indexed by a set of variables, such that, whenever $P \mathrel{\mathcal{R}^{\mathcal{E}}} Q$ the following hold:
\begin{itemize}
\item If $\left\{x,y\right\} \cap \mathcal{E} = \emptyset$ then $P\sub{x}{y}  \mathrel{\mathcal{R}^{\mathcal{E}}} Q\sub{x}{y}$.
\item If $P \lts{\alpha} P'$ then there exists $Q'$ such that $Q \lts{\alpha} Q'$ and $P'  \mathrel{\mathcal{R}^{\mathcal{E}}} Q'$.
\item If $P \lts{\co{a}(x)} P'$ then there exists $Q'$ such that $Q \lts{\co{a}(x)} Q'$ and $P'  \mathrel{\mathcal{R}^{\mathcal{E}, x}} Q'$.
\item If $P \lts{a(x)} P'$ then there exists $Q'$ such that $Q \lts{a(x)} Q'$ and, for all $x$, $P'  \mathrel{\mathcal{R}^{\mathcal{E}}} Q'$.
\end{itemize}
Intermediate bisimilarity $\isim$ is the greatest intermediate bisimulation.
\end{defi}
Intermediate bisimilarity, a secondary contribution of this paper, defined above, sits between open bisimilarity, late equivalence and open barbed bisimilarity.
Intermediate bisimilarity is a congruence, hence is sound with respect to open barbed bisimilarity.
Strictness of this inclusion follows since
$\mathopen{\nu k.\co{a}k.}\left( a(x).\tau + a(x) \right)$
and 
$\mathopen{\nu k.\co{a}k.}\left( a(x).\match{x = k}\tau + a(x).\tau + a(x) \right)$
are distinguished by intermediate bisimilarity, as witnessed by the following strategy.

\qquad
\xymatrix@R=5pt{
&\mathopen{\nu k.\co{a}k.}\left( a(x).\tau + a(x) \right)
\ar^{\co{a}(k)}[d] 
&& \not\isim &
\mathopen{\nu k.\co{a}k.}\left( a(x).\tau + a(x) \right)
\ar^{\co{a}(k)}[d] 
\\
&a(x).\tau + a(x)  
\ar@/_/^{a(x)}[dl]\ar@/^/_{a(x)}[dr]
&& \not\isim^{k} &
 a(x).\match{x = k}\tau  +  a(x).\tau \!+\! a(x) 
\ar^{a(x)}[d]
\\
\tau && 0 
&&   \match{x = k}\tau
}

Observe that, there are two cases to check at this point: either we apply substitution $\sub{x}{k}$, in which case we have $0 \not\isim^k \match{k = k}\tau\sub{x}{k}$;
or we apply any other substitution for $x$, say $\sigma$, in which case $\match{x\sigma = k}\tau$ is deadlocked, hence $\tau \not\isim^k \match{x\sigma = k}\tau$.
In contrast, as remarked  previously, these processes are open barbed bisimilar.

Intermediate bisimilarity is strictly coarser than open bisimilarity.
To see why, observe the following processes are equivalent according to intermediate bisimilarity.
\[
\mathopen{\nu k.\co{a}k.a(x).}\left( \tau + \tau.\tau + \tau.\match{x = k}\tau \right)
\isim
\mathopen{\nu k.\co{a}k.a(x).}\left( \tau + \tau.\tau \right)
\]
In contrast, for open bisimilarity, there is a distinguishing strategy for the same pair of processes, as witnessed by the following formula in \OM.
\[
\mathopen{\nu k.\co{a}k.a(x).}\left( \tau + \tau.\tau + \tau.\match{x = k}\tau \right)
 \vDash \diam{a(k)}\diam{a(x)}\diam{\tau}\left( x = k \yields \diam{\tau}\ttt \wedge \boxm{\tau}\left( x = k \right) \right)
\]
The difference is, when constructing an open bisimulation, we can proceed with the first $\tau$ transition without deciding whether $x = k$ or $x \not= k$.
In contrast, intermediate bisimilarity forces this decision immediately after $x$ is input.

It is important to note that we are not advocating that intermediate bisimilarity should be used in preference to open bisimilarity.
What we are emphasising here is that open bisimilarity does not hold a canonical status as a bisimilarity congruence sound with respect to late bisimilarity.
Indeed, there is a spectrum bisimilarities between open bisimilarity and open barbed bisimilarity.

\begin{figure}[b]
\xymatrix@C=-33pt{
                                &\txt{early bisimilarity \\ classical $\FM$~\cite{Milner93}}  &  \\
\txt{late bisimilarity \\ classical \OM,~Fig.~\ref{figure:classicalOM}}\ar[ur] &                                         & \txt{barbed equivalence~\cite{Sangiorgi1992} \\ early equivalence }\ar[ul]            & \ar@{=}[dddlll] \\
                                &\mbox{late equivalence~\cite{Milner92}}\ar[ul]\ar[ur]    &                                             & \ar[ul]\txt{open barbed bisimilarity~\cite{Sangiorgi2001} 
 \\ intuitionistic $\FM$~\cite{Horne2018}} \\
           \textit{classical}   &                                         & \txt{intermediate bisimilarity 
}\ar[ul]\ar[ur]  & \\
                                &  \textit{non-classical}                 && \txt{open bisimilarity~\cite{sangiorgi96acta} \\ intuitionistic \OM,~Fig.~\ref{figure:om}}\ar[ul]              
}
\caption{The line between classical and non-classical notions of bisimilarity.}\label{fig:non-classical}
\end{figure}

A picture of part of the spectrum surrounding open bisimilarity is provided in Fig.~\ref{fig:non-classical}.
To complete the picture in Fig.~\ref{fig:non-classical}, note that related work~\cite{Horne2018} introduced a modal logic characterising open barbed bisimilarity called intuitionistic $\FM$ --
the intuitionistic counterpart to a classical modal logic characterising early bisimilarity.
That paper emphasises the merits of open barbed bisimilarity due to its more objective definition, and, more importantly still, its coarser granularity suitable for verifying privacy properties.  Open barbed bisimilarity can be used to verify properties of protocols that make use of else branches to maintain the privacy of honest participants; whereas open bisimilarity fails to verify such scenarios, instead discovering spurious attacks. This is due to the intuitionistic nature of open bisimilarity, which, as we have seen, assumes the absence of the law of excluded middle everywhere; whereas open barbed bisimilarity verifies more properties since it induces the law of excluded middle for private names only. Related work~\cite{arXiv}, elaborates on this perspective, making a case for why variants of open barbed bisimilarity and $\FM$ are, respectively, the notions of bisimilarity congruence and intuitionistic modal logic suited to the applied $\pi$-calculus; rather than open bisimilarity and $\OM$.

\subsection{Related work: an alternative logic formalised in Nominal Isabelle}
Parrow et al.~\cite{Parrow15}\ provided a general proof of
the soundness and completeness of logical equivalence for various modal logics
with respect to corresponding bisimulations.
The proof is parametric on properties of substitutions, which can be instantiated
for a range of bisimulations. Moreover, their proof is
mechanised using Nominal Isabelle.
The conference version~\cite{Parrow15} sketches how
to instantiate their abstract framework for open bisimilarity for a fragment of the $\pi$-calculus without input prefixes.
A forthcoming journal version~\cite{Parrow19} generalises their methodology such that open bisimilarity with input prefixes may also be handled.

Stylistically, our intuitionistic modal logic is quite different from an instantiation of the abstract framework of Parrow et al.\ for open bisimilarity.
Their framework, is classical and works by syntactically restricting ``effect'' modalities in formulae, depending on the type of bisimulation. Their effects represent substitutions that reach worlds permitted by the type of bisimulation. 
In contrast, the modalities of the intuitionistic modal logic $\mathcal{OM}$ in this paper are syntactically closer to long established modalities for the $\pi$-calculus~\cite{Milner93}; differing instead in their intuitionistic interpretation.
An explanation for the stylistic differences is that for every intuitionistic logic, such as the intuitionistic modal logic in this work,
there should be a corresponding classical modal logic based on an underlying Kripke semantics.
Such a Kripke semantics would reflect the accessible worlds, as achieved by the syntactically restricted effect modalities in the abstract classical framework instantiated for open bisimilarity.

\section{Mechanising the Soundness Proof in Abella}\label{sec:abella}

Abella~\cite{Baelde14} is a proof assistant based on intuitionistic logic that supports
both inductive and coinductive reasoning over logical specifications of operational semantics
for languages that contain binding structures, such as the $\pi$-calculus. 
In particular, Abella is well-suited for reasoning involving operational semantics
specified in the higher-order logic programming language \lProlog~\cite{lprolog}. 
The formalisation of the modal logic \OM\ in this section is built on top of
existing work on the formalisation of the $\pi$-calculus and bisimulation
based on the higher-order abstract syntax (HOAS) approach~\cite{Bedwyr07,TiuMil10,Baelde14}. 
We present the coinductive definition of open bisimilarity (Section~\ref{sec:bisimdef}) and
the semantics of the modal logic \OM\ (Section~\ref{sec:abella:sound}) formalised in Abella,
leading up to our mechanised proof of the soundness theorem (Theorem~\ref{theorem:sound}).

Interestingly, the proof of soundness (Theorem~\ref{theorem:sound}) is quite abstract since it can be proven without defining a specific language of process terms and their labelled transition system rules,
since the proof only looks at the labels and makes the implicit assumption that transitions satisfy monotonicity.
Thus although it is not required for the main theorems of this paper, we none-the-less, for a self-contained presentation also recall an established \lProlog\ specification of the $\pi$-calculus at the end of this section (Section~\ref{sec:pispec}), which can be used as the basis of tooling.

\begin{figure}
\begin{lstlisting}[numbers=left, numberstyle=\tiny\sf\color{gray}, mathescape=true]
Specification "finite-pic". % load the finite pi-calc. spec. in $\color{gray}\text{Fig.\,\ref{fig:fpic}}$

CoDefine bisim : p -> p -> prop  % open bisimulation
by bisim P Q
  := (forall A P1, {one  P  A     P1} -> exists Q1, {one  Q  A     Q1} /\      bisim P1 Q1)
  /\ (forall X M, {oneb P (dn X) M} -> exists N, {oneb Q (dn X) N} /\ forall z, bisim (M z) (N z))
  /\ (forall X M, {oneb P (up X) M} -> exists N, {oneb Q (up X) N} /\ nabla z, bisim (M z) (N z))
  /\ (forall A Q1, {one  Q  A     Q1} -> exists P1, {one  P  A     P1} /\      bisim Q1 P1)
  /\ (forall X N, {oneb Q (dn X) N} -> exists M, {oneb P (dn X) M} /\ forall z, bisim (N z) (M z))
  /\ (forall X N, {oneb Q (up X) N} -> exists M, {oneb P (up X) M} /\ nabla z, bisim (N z) (M z)).

Kind o'  type.              % syntax of the modal logic
Type tt, ff          o'.
Type conj, disj, impl         o' -> o' -> o'.
Type mat               n -> n -> o'.
Type boxAct, diaAct            a -> o' -> o'.
Type boxOut, diaOut, boxIn, diaIn      n -> (n -> o') -> o'.

Define sat : p -> o' -> prop  % semantics of the modal logic
by sat P tt
 ; sat P (conj A B) := sat P A /\ sat P B
 ; sat P (disj A B) := sat P A \/ sat P B
 ; sat P (impl A B) := sat P A -> sat P B  
 ; sat P (mat X Y)   := X = Y
 ; sat P (boxAct X A) := forall P1, {one P X P1} -> sat P1 A
 ; sat P (diaAct X A) := exists P1, {one P X P1} /\ sat P1 A
 ; sat P (boxOut X A) := forall Q, {oneb P (up X) Q} ->  nabla z, sat (Q z) (A z)
 ; sat P (diaOut X A) := exists Q, {oneb P (up X) Q} /\  nabla z, sat (Q z) (A z)
 % basic input modality (see $\text{\textcolor{gray}{\textrm{Section~\ref{sec:related:nabla}}}}$ for related discussion)
 ; sat P (boxIn X A) := forall Q, {oneb P (dn X) Q} ->  forall z, sat (Q z) (A z)
 % late input modality  (see $\text{\textcolor{gray}{\textrm{Section~\ref{sec:related:nabla}}}}$ for related discussion)
 ; sat P (diaIn X A) := exists Q, {oneb P (dn X) Q} /\  forall z, sat (Q z) (A z).
 \end{lstlisting}
\caption{A coinductive definition of open bisimulation and
       an inductive definition of the modal logic \OM\ in Abella.}
\label{fig:BisimSatDefs}
\end{figure}

\subsection{Coinductive definition of open bisimulation in Abella's reasoning logic}
\label{sec:bisimdef}
Open bisimulation relation \lstinline|bisim| is coinductively defined in Fig.~\ref{fig:BisimSatDefs}.
The relation \lstinline|bisim| is an Abella encoding of
the open bisimulation relation $\mathcal{R}$ in Definition~\ref{def:bisim} from Section~\ref{sec:openbisim}.
Lines 5, 6, and 7 correspond to the latter three of the four bullet items in Definition~\ref{def:bisim},
which state the closure property under every pairwise bisimulation step
where \lstinline|P| leads and \lstinline|Q| follows.
Lines 8, 9, and 10 are symmetric cases
where \lstinline|Q| leads and \lstinline|P| follows.
Curly braces (e.g., \lstinline|{one P A P1}|) are used for referring to
the object logic proposition (i.e., \lProlog\ proposition) from the reasoning logic of Abella.
The \lProlog\ relation \lstinline|one : p->a->p->o|
(see Section~\ref{sec:pispec} for further details), when applied to three arguments,
becomes an object logic proposition \lstinline|one P A P1 : o|.
In order refer to such \lProlog\ propositions from Abella's reasoning logic,
we use curly braces to convert a \lProlog\ proposition (\lstinline|o|)
into a reasoning logic proposition (\lstinline|prop|).
For instance, \lstinline|{one P A P1} :  prop|.
Abella's reasoning logic is richer than the object logic.
It supports coinductive definitions, which we used to define \lstinline|bisim|.
It also supports nominal quantification (\lstinline|nabla|), which will be discussed shortly,
in addition to universal (\lstinline|forall|) and existential (\lstinline|exists|) quantifications.

The first bullet item in Definition~\ref{def:bisim} states that open bisimulation must be closed under
all substitutions that respect the history.
In the definition of \lstinline|bisim|, Abella guarantees this closure property under respectful substitutions for free.
Let us first demonstrate how histories are being handled in the relation \lstinline|bisim|,
in order to explain how the closure property under respectful substitutions is ensured in Abella.
Consider a trivial bisimulation over identical processes, illustrated using both Abella and mathematical notations as follows:\footnote{Here, mathematical notations of process terms are used on the left side, instead of the actual \lProlog\ embeddings, as well as on the right side. The embeddings of process terms are provide in Section~\ref{sec:pispec}.}

\[
\infer{ \hspace{-16ex}\text{\scriptsize\lstinline|forall|\lstinline|x,|~
  \texttt{bisim ({\normalsize$\,\nu z.\co{x}{z}.x(y).0\,$})
		({\normalsize$\,\nu z.\co{x}{z}.x(y).0\,$})}\hspace{-6ex}}^{\phantom{G^G}} 
}{ \quad \infer{ \text{\scriptsize\lstinline|forall|\lstinline|x,nabla|\lstinline|z,|~\texttt{bisim ({\,\normalsize$x(y).0\,$}) ({\,\normalsize$x(y).0\,$})}}_{\phantom{G}}^{\phantom{G^G}} 
	}{ \text{\scriptsize\lstinline|forall|\lstinline|x,nabla|\lstinline|z,forall|\lstinline|y,|~\texttt{bisim} {\normalsize$~0$} {\normalsize$~0$}} \qquad
    & ~\vdots\quad
   }
 & \qquad\qquad\vdots\qquad\quad
}_{\phantom{G_G}}
\infer{ \nu z.\co{x}{z}.x(y).0 ~\sim^{\,x^i}~ \nu z.\co{x}{z}.x(y).0
}{ \infer{ x(y).0 ~\sim^{\,x^i\cdot z^o}~ x(y).0
   }{ 0 ~\sim^{\,x^i\cdot z^o\cdot y^i}~ 0 
    & ~~\vdots~
   }
 & \quad\vdots~~
}
\]

Even for identical processes without nondeterministic constructs,
the bisimulation tree has at least two branches for each node
because either one of the two sides may take a leading step to be followed by the other side.
Here, let us focus on the leftmost branches where the left process leads.
The environment of quantified variables for the Abella relation \lstinline|bisim| grows after each bisimulation step.
Growing the environment exactly corresponds to growing the history.
The bound output step extends the environment with \lstinline|nabla|\lstinline|z| in Abella,
which corresponds to extending the history with $z^o$.
The input step extends the environment with \lstinline|forall|\lstinline|y| in Abella,
which corresponds to extending the history with $y^i$.
These quantified variables come from the definition of \lstinline|bisim| in Fig.~\ref{fig:BisimSatDefs}, more specifically, from lines 6 and 7.

Recall the definition of respectful substitution (Definition~\ref{def:respects}) from Section~\ref{sec:direct}.
An input variable in the history adds no restriction to the respectfulness of a substitution.
An extruded private name in the history adds a restriction such that respectful substitutions should not unify
the output variable with any variable that precedes the output variable.
The nabla quantifier (\lstinline|nabla|) in Abella coincides with such a notion of restriction.
Nabla quantified variables are guaranteed to be fresh names with respect to all the previously introduced names.
For instance, consider the environment
{\small\lstinline|forall|\lstinline|x,nabla|\lstinline|z,forall|\lstinline|y,|$\cdots$}.
Abella ensures that \lstinline|z| cannot occur free in \lstinline|x|, hence,
\lstinline|x| cannot be unified with \lstinline|z|; however, \lstinline|y|
can be unified with \lstinline|z| because \lstinline|y| is introduced after \lstinline|z|.

Intuitively, universal quantification represents all possible substitutions over universally quantified variables.
For example, consider \lstinline|forall x, pred x.|
Proving this in Abella means that the predicate \lstinline|pred| holds for all possible substitutions over \lstinline|x|.
Together with nabla quantification, the notion of all possible respectful substitutions can be represented by the environment of quantified variables in Abella.
In summary, the list of universal and nabla quantified variables before the \lstinline|bisim| relation in Abella not only transcribes the history but also represents all possible respectful substitutions.

\subsection{Embedding of \OM\ in Abella and the soundness proof}
\label{sec:abella:sound}

The latter part of Fig.~\ref{fig:BisimSatDefs} is an embedding of the syntax and semantics
of \OM\ introduced earlier in Section~\ref{sec:om}. Recall the stylistic difference between Abella and
mathematical notations for the process syntax~-- prefixes of free actions, bound output actions,
and input actions are defined as three different syntactic constructs in the Abella definitions
(see Fig.~\ref{fig:BisimSatDefs}). There are similar stylistic difference regarding \OM\ formulae
in the Abella embedding (Fig.~\ref{fig:BisimSatDefs}) and the notation in Section~\ref{sec:om}.
There are three formulae constructs for each kind of modality. For instance,
\lstinline|boxAct|, \lstinline|boxOut|, and \lstinline|boxIn|$~$ are the three different syntactic constructs
of the box modality for free actions, bound output actions, and input actions, respectively.
Similarly, there are three constructs for the diamond modality.\footnote{In the Abella proof scripts, 
\lstinline|boxAct|, \lstinline|boxOut|, and \lstinline|boxIn|$~$ are represented using
keywords \texttt{boxAct}, \texttt{boxOut} and \texttt{boxIn}, respectively.}

Recall that histories on the bisimulation relation are transcribed as universal and
nabla quantified variables in Abella and that closure under respectful substitutions
holds for free in Abella. Similarly, histories in the semantics of \OM\ are handled in exactly
the same manner in Abella and enjoy the closure properties regarding respectful substitutions.
For example, \lstinline|forall|\lstinline|x,nabla|\lstinline|z,forall|\lstinline|y, sat (P x y z) $\phi$| $~$
corresponds to 
$ \forall \sigma\respecting x^i \cdot z^o\cdot y^i,~P(x,y,z) \models^{x^i\cdot z^o\cdot y^i} \phi $.
The relation \lstinline|sat| in Fig.~\ref{fig:BisimSatDefs} is an embedding of
\OM\ ($\models$) in Abella. There is no explicit handing of substitutions in the semantics of
the definition of the \lstinline|sat| relation  because they are handled by Abella automatically.

We mechanised the proof of soundness of open bisimilarity with respect to \OM\ by 
proving the following theorem.
\begin{lstlisting}
Theorem bisim_sat : forall P Q F, form F ->
                          bisim P Q -> ((sat P F -> sat Q F) /\ (sat Q F -> sat P F)).
\end{lstlisting}
In the above, \lstinline{form: o'->prop} is an inductive predicate for well-formed \OM\ formulae, defined as follows:
\begin{lstlisting}
Define form : o' -> prop
by form tt
 ; form ff
 ; form (conj A B) := form A /\ form B
 ; form (disj A B) := form A /\ form B
 ; form (impl A B) := form A /\ form B
 ; form (mat X Y) := form A
 ; form (boxAct X A) := form A
 ; form (diaAct X A) := form A
 ; form (boxOut X A) := forall w, form (A w)
 ; form (diaOut X A) := forall w, form (A w)
 ; form (boxIn X A) := forall w, form (A w)
 ; form (diaIn X A) := forall w, form (A w).
\end{lstlisting}
The predicate \lstinline{form} is a trick used to guide the induction in Abella by the structure of formulae (instead of by the structure of \lstinline{sat}, which is not stratified~\cite{Momigliano2004}, due to the presence of implication).
Thereby the proof of theorem \lstinline{bisim_sat} below is established by induction on the structure of the modal formulae and by case analyses on the definition of the satisfiability relation \lstinline{sat}. 
\begin{lstlisting}
Theorem soundness: forall P Q, 
  bisim P Q -> forall F, form F -> (sat P F -> sat Q F) /\ (sat Q F -> sat P F).
\end{lstlisting}
The soundness theorem  is a simple corollary of \lstinline{bisim_sat}, since \lstinline{form F} is a tautology.
Due to the adequacy of this embedding, this concludes the mechanisation of the soundness of open bisimilarity with respect to $\OM$ (Theorem~\ref{theorem:sound}).

We have not yet discussed the \lProlog\ embedding of the labelled transition system
because specific details of the transition system do not affect the soundness property.
During the proof of \lstinline|bisim_sat|,
the transition relations (\lstinline|one| and \lstinline|oneb|) are never inspected,
nor are processes inspected.
The only specific details of the labelled transition system, which are evident in the definitions of
\lstinline|bisim| and \lstinline|sat|, are the transition labels \lstinline|up| and \lstinline|dn|.
Thus, the minimal definitions that are required in the specification \lstinline|finite-pic| of
the soundness proof are:
\begin{lstlisting}
kind n  type. % names
kind p  type. % processes
kind a  type. % actions (transition labels)
type up, dn  n -> n -> a.

type one   p ->       a  ->          -> o.  % one step  free transition
type oneb  p -> (n -> a) -> (n -> p) -> o.    % one step bound transition
one  P A Q :- $~~\cdots~\cdots$ .  $\qquad$ % arbitrary transition definition suffice
oneb P A Q :- $~~\cdots~\cdots$ .  $\qquad$ % arbitrary transition definition suffice
\end{lstlisting}
Additional details of the processes syntax and the transition relations
are irrelevant to the soundness property.
Recall that both the bisimilarity (\lstinline|bisim|) and modal logic satisfaction (\lstinline|sat|)
is defined in terms of single step transition relations \lstinline|one| and \lstinline|oneb|.
The soundness property could be understood in term of the question of whether the use of transition relations
in \lstinline|bisim| match well with the use of those relations in \lstinline|sat|. 
Regardless of how we specify the process syntax and the transition relations,
\lstinline|bisim_sat| continues to hold in Abella without any modification to our proof script.
This indicates that our soundness result (and also completeness, for similar reasons)
is independent of the terms of the processes calculus, relying only on
the form of the transition labels and monotonicity of the labelled transition system.
	Note, the style of embedding for the labelled transition system,
	which relies on Abella's logical quantifiers for implicitly embedding histories,
	already assumes monotonicity of labelled transitions (Proposition~\ref{prop:monotonicity}) due to the intuitionistic nature of Abella.

\begin{figure}
  % add line numbers to be coherent with Fig 5
  \begin{lrbox}{\mybox}
    \begin{minipage}{1.1\linewidth}
\begin{lstlisting}[numbers=left, numberstyle=\tiny\sf\color{gray}, mathescape=true]
sig finite-pic. % file: finite-pic.sig

kind n  type. % names

kind p  type. % processes
type null         p.                 % deadlock 
type taup         p -> p.             % progress action
type plus, par    p -> p -> p.         % choice, par
type match, out   n -> n -> p -> p.     % match, output action
type in           n -> (n -> p) -> p.   % input action
type nu           (n -> p) -> p.       % nu

kind a  type. % actions (transition labels)
type tau          a.
type up, dn       n -> n -> a.

type one     p ->      a  ->      p  -> o.  % one step  free transition
type oneb    p -> (n -> a) -> (n -> p) -> o.  % one step bound transition


module finite-pic. % file: finite-pic.mod

oneb (in X M) (dn X) M.      % bound input
one (out X Y P) (up X Y) P.  % free output
one (taup P) tau P.          % tau
% match prefix
one (match X X P) A Q :- one  P A Q.     oneb (match X X P) A M :- oneb P A M.
% sum
one (plus P Q) A R :- one  P A R.        oneb (plus P Q) A M :- oneb P A M.
one (plus P Q) A R :- one  Q A R.        oneb (plus P Q) A M :- oneb Q A M.
% par
one (par P Q) A (par P1 Q) :- one P A P1.  oneb (par P Q) A (z\par (M z) Q) :- oneb P A M.
one (par P Q) A (par P Q1) :- one Q A Q1.  oneb (par P Q) A (z\par P (N z)) :- oneb Q A N.
% restriction
one  (nu z\P z) A (nu z\Q z)     :- pi z\ one  (P z) A (Q z).
oneb (nu z\P z) A (y\nu z\Q z y) :- pi z\ oneb (P z) A (y\Q x z).
% open (bound output)
oneb (nu x\P x) (up X) Q :- pi y\ one (P y) (up X y) (Q y).
% close
one (par P Q) tau (nu z\par (M z) (N z)) :- oneb P (dn X) M, oneb Q (up X) N.
one (par P Q) tau (nu z\par (M z) (N z)) :- oneb P (up X) M, oneb Q (dn X) N.
% comm (interaction)
one (par P Q) tau (par (M Y) T) :- oneb P (dn X) M, one Q (up X Y) T.
one (par P Q) tau (par R (M Y)) :- oneb Q (dn X) M, one P (up X Y) R.
\end{lstlisting}
    \end{minipage}
  \end{lrbox}
  \scalebox{0.909}{\usebox{\mybox}}
\caption{\lProlog\ specification of the syntax and transitions of the $\pi$-calculus.
          (Adopted from one of the examples distributed with Abella.
          The adequacy for this \lProlog\ embedding has been justified
          by Tiu and Miller \cite{TiuMil10}.)}
\label{fig:fpic}
\end{figure}

\subsection{Specification of the syntax and labelled transition systems of the $\pi$-calculus in \lProlog}
\label{sec:pispec}
Despite the abstract nature of our soundness result, discussed above, in order to check specific examples of bisimilarity and satisfaction in Abella, an embedding of the syntax and labelled transition system of the $\pi$-calculus in Fig.~\ref{figure:pi} is required.
Fig.~\ref{fig:fpic} is a transcription of the syntactic constructs and transition rules into \lProlog.
This embedding of the $\pi$-calculus labelled transition system and bisimulation
originates form the work of Tiu and Miller \cite{TiuMil10}, where they provide
rigorous adequacy result for their logical embedding, which has an obvious direct
transcription into \lProlog. By adequacy we mean that standard definitions and the logical embedding prove the same theorems.

Each syntactic category is declared as a type (e.g., \lstinline|n|, \lstinline|p|, and \lstinline|a|)
using the \lstinline|kind| keyword. Syntactic constructs are defined as constants in
\lProlog\ where some of which may require multiple arguments to construct the desired syntactic category.
For instance, \lstinline|plus| needs two process arguments to construct a process, as its type
(\lstinline|p-> p-> p|) suggests. 
The table below summarises the process syntax in \lProlog\ and the notations used in the previous sections.
\begin{table}[h!]
\begin{tabular}{lll}
\lProlog\ syntax & mathematical notation \\
\hline
\lstinline|null|                      & \quad $0$ \\
\lstinline|nu x\P x| \quad$~~$       
(or$~$ \lstinline|nu P|)              & \quad $\nu z.P$ & \lstinline|(P z)| corresponds to $P$ \\
\lstinline|taup P|                    & \quad $\tau.P$ \\
\lstinline|out x z P|                 & \quad$\co{x}{z}.P$ \\
\lstinline|in x z\P z| $~~$
(or$~$ \lstinline|in x P|)            & \quad $x(z).P$ & \lstinline|(P z)| corresponds to $P$ \\
\lstinline|match x y P|               & \quad $[x=y]P$ \\
\lstinline|par P Q|                   & \quad $P \cpar Q$ \\
\lstinline|plus P Q|                  & \quad $P + Q$ \\
\end{tabular}
\end{table}

Fig.~\ref{fig:fpic} has few stylistic differences from Fig.~\ref{figure:pi}.
Firstly, distinct constants are used for actions and their related processes
(e.g., a \lstinline|tau| action for \lstinline|taup| prefixed process) because the constants
cannot be overloaded as in mathematical notation. Secondly, the action prefix
$\pi.P$ in Fig.~\ref{figure:pi}, where $\pi$ ranges over several different types of action
(progress, free out, and input), is transcribed as three distinct process syntactic constructs
(\lstinline|taup|, \lstinline|out|, and \lstinline|in|).
Thirdly, free and bound actions are distinguished by their types instead of
using different notations ($\co{x}{z}$ and $\oprivate{x}{z}$) as in Fig.~\ref{figure:pi}.
That is, bound actions (e.g., \lstinline|up x : n->a|) are partially applied free actions
(e.g., \lstinline|up x z : a|) using the same constants.
Fourthly, two different sets of transition relations are defined:
\lstinline|one| relating a process (\lstinline|p|) to another process (\lstinline|p|) via a free action (\lstinline|a|) and
\lstinline|oneb| relating a process (\lstinline|p|) with a bound process (\lstinline|n->p|) via a bound action (\lstinline|n->a|).

One advantage of using \lProlog~\cite{lprolog} is that we can rely on its native support for a variant of HOAS,
known as \emph{$\lambda$-tree syntax}~\cite{miller99surveys},
for handling bound variables and $\alpha\beta\eta$-equivalence automatically.
For instance, consider the rule for name extrusion
from both Fig.~\ref{fig:fpic}
and
Fig.~\ref{figure:pi}:
\begin{center}
\begin{tabular}{ll}
$\begin{array}{ll}
{\small\texttt{
	oneb (nu z\textbackslash P z) (up X) Q}} ~\text{~\lstinline|:-|} \\ \qquad
{\small\texttt{pi z\textbackslash one (P z) (up X z) (Q z)}} \\
~
\end{array}$
& \qquad
\infer[x \not= z]{
\nu z. P \lts{\oprivate{x}{z}} Q
}{
P \lts{\co{x}z} Q
}
\end{tabular}
\end{center}
There is no need to explicitly state and keep track of the side conditions
such as $x\neq z$ and $x\notin \mathop{\mathrm{n}}(\pi)$ in \lProlog\ definitions.
For example, consider $~${\small\texttt{pi z\textbackslash one (P z) (up X z) (Q z)}} 
from above. Here, it is guaranteed that \texttt{z} does not to occur free in
the logic variables \texttt{P}, \texttt{X}, and \texttt{Q}. This guarantee
comes from the scoping of variables: the scopes of \texttt{P}, \texttt{X}, and \texttt{Q}
go beyond the scope of \texttt{z}, which is limited only to $~${\small\texttt{one (P z) (up X z) (Q z)}}.
Had \texttt{z} freely occurred in any of \texttt{P}, \texttt{X}, or \texttt{Q},
the scope of \texttt{z} would have been violated. Hence, \texttt{X} cannot be unified with \texttt{z}.

\section{Distinguishing formulae generation algorithm implementation}%
\label{sec:dfalgo}
\begin{figure}
\includegraphics[width=\linewidth]{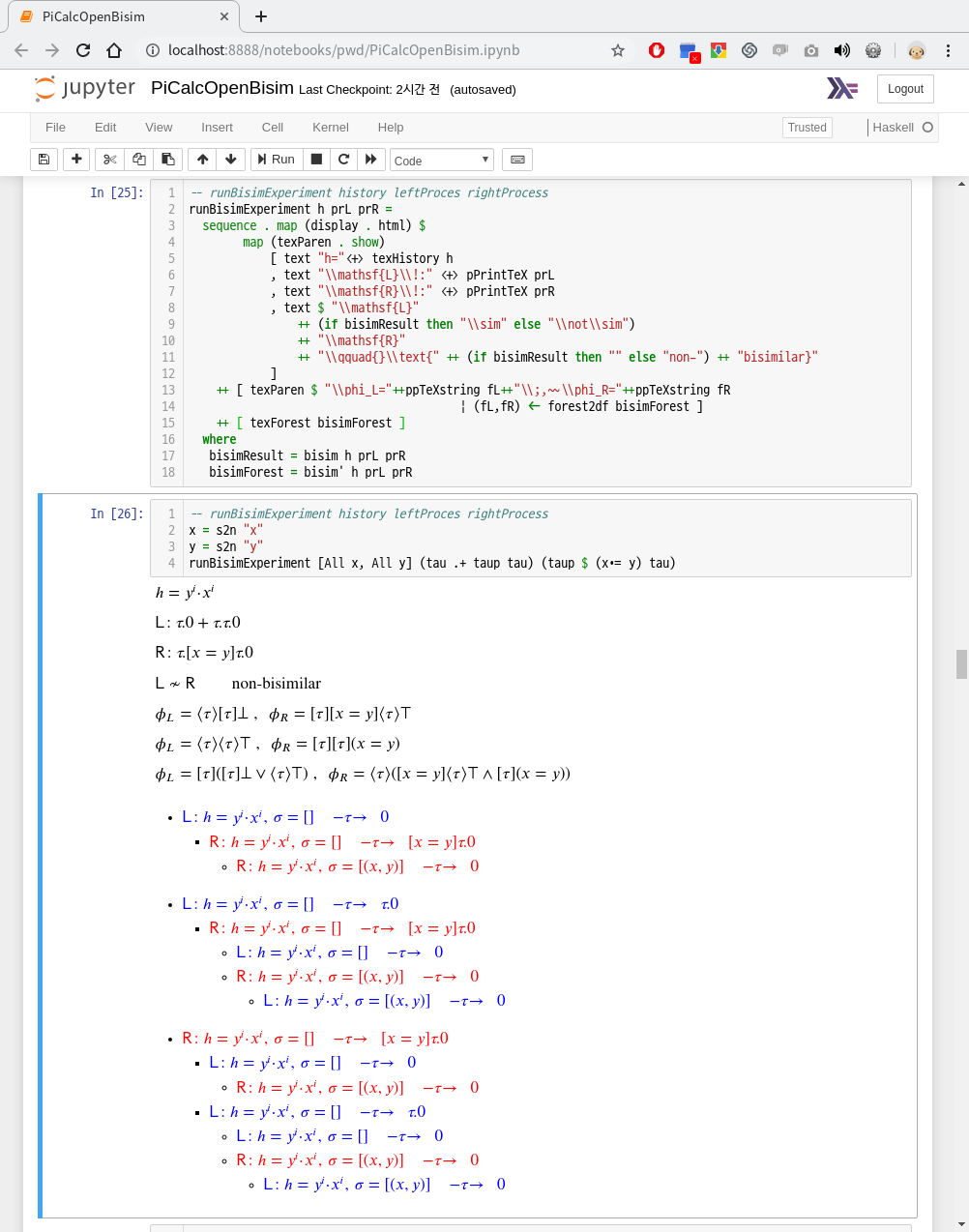}
\caption{Screenshot of our distinguishing formulae generation algorithm running in Jupyter.}
\label{fig:ihaskell}
\end{figure}

\lstset{language=Haskell,basicstyle=\ttfamily\footnotesize}

Our completeness proof in Section~\ref{sec:complete} is constructive in the sense
that it follows the structure of an algorithm (Section~\ref{sec:complete:algo})
to build a pair of formulae for a pair of distinguishable processes. That is,
one can find distinguishing formulae for any given pair of processes when they are
actually distinguishable, guided by the steps described in our completeness proof
(Section~\ref{sec:complete:proof}). These steps can be automated by writing a program
that implements this algorithm. We first implemented the distinguishing formulae 
generation algorithm using Haskell~\cite{AhnHorTiu17genwit}, accompanying
the conference publication~\cite{Ahn2017} of our work. Here, we provide pointers
to our current implementation and describe continuing work to make it more accessible
to those who are not accustomed to Haskell development tools including the GHC compiler.

Our current implementation is available online from a public GitHub repository\footnote{\small\url{https://github.com/kyagrd/ihaskell-picalc}}. This repository contains
an example notebook and some utility shell scripts, which utilize a Docker container
published on DockerHub to run the example notebook. All the necessary software
dependencies, including Jupyter and Haskell, to run our implementation is contained
in a Docker image so that it does not interfere with your own system. Anyone with
access to an Internet-connected machine with a properly working Docker system
can easily run our implementation with a single shell script command, although
it would initially require some download time and disk space for a sizable ($<$10GB)
Docker image to start running.

Figure~\ref{fig:ihaskell} is a screenshot of a web browser connected to
a Jupyter server running on the same machine (i.e., localhost).
The Haskell source code documented in our technical report~\cite{AhnHorTiu17genwit}
is executed via IHaskell, a Haskell language kernel for the Jupyter notebook environment.
Some additional features are the enhanced output
exploiting Jupyter notebook's ability to render HTML and also some LaTeX via MathJax.
In addition to the plain text output, which you can still use inside the Jupyter notebook,
we provide LaTeX output that looks the same as the notation used in this article.

In Figure~\ref{fig:ihaskell}, we define the function \lstinline|runBisimExperiment|
and demonstrate it used on a pair of non-bisimilar processes, discussed previously in Section~\ref{eg:law}.
We are providing the implementation as a Haskell library so that one can build
programs that automate tasks related to bisimulation.
The function \lstinline|runBisimExperiment| is defined in terms of
more primitive definitions provided in our implementation. This function is applied
to three arguments: the initial history, the left processes, and the right process.
The free variables of the two processes must be closed by the initial history,
either as input (using \lstinline|All|) as output (using \lstinline|Nab|).
In the example run, we closed the two free variables \lstinline|x| and \lstinline{y}
as inputs in the initial history \lstinline|[All x, All y]|.
To provide processes, one should build them as Haskell values representing
abstract syntax trees, whose data type is defined as follows:
\begin{lstlisting}[language=Haskell]
    import GHC.Generics (Generic)
    import Unbound.Generics.LocallyNameless -- using unbound-generics library
    
    type Nm = Name Tm
    newtype Tm = Var Nm deriving (Eq, Ord, Show, Generic)
    
    data Pr  = Null | TauP Pr | Out Tm Tm Pr | In Tm PrB | Match Tm Tm Pr
             | Plus Pr Pr | Par Pr Pr | Nu PrB  deriving (Eq, Ord, Show, Generic)
    type PrB = Bind Nm Pr
    
    instance Eq PrB where (==) = aeq;  instance Ord PrB where compare = acompare
    instance Alpha Tm;  instance Alpha Pr;  instance Subst Tm Pr
\end{lstlisting}
For instance,
the procsseses $\tau.0 + \tau.\tau.0$ is written as
\lstinline|(Plus (TauP Null) (TauP(TauP Null)))|
and the process $\tau.[x=y]\tau.0$ 
is written as \lstinline|(TauP(Match (Var x) (Var y) (TauP Null)))|
using the data type definition above.
% which is not unreadable but rather verbose.
We additionally provide shorthand definitions below to reduce keyboard strokes
and to make it look closer to the notations used in this article.
\begin{lstlisting}[language=Haskell]
    (.\) = bind; infixr 1 .\ ;   (.+) = Plus; infixl 6 .+ ;  (.|)  = Par; infixl 5 .|
    o = Null ;  taup = TauP ; out x y = Out(Var x)(Var y) ;  inp = In . Var ;  nu = Nu
    x .= y = Match (Var x) (Var y) ;      tau = TauP Null ;  tautau = TauP (TauP Null)
\end{lstlisting}
Using these shorthand definitions, we can write $\tau.0 + \tau.\tau.0$ and $\tau.[x=y]\tau.0$
as \lstinline|(tau .+ taup tau)| and \lstinline|(taup $\texttt{\$}$ (x.=y) tau)|.

As a result of running the function \lstinline|runBisimExperiment|,
it displays the following:
\begin{itemize}
\item the initial history and the two processes,
\item the result of bisimilarity test,
\item a pair of distinguishing formulae (for non-bisimilar processes), and
\item a set of trees consisting of all transitions in a distinguishing strategy.
\end{itemize}
The first two items are self explanatory from its rendered output.
Let us explain few additional details on the last two items above.

Our implementation handles a subset of \OM-formulae (Defintion~\ref{def:OMsyntax})
that is sufficient to generate distinguishing formulae for any pair of non-bisimilar processes.
The distinguishing formulae constructed during our completeness proof
contain a limited form of implication $(x=y)\supset\phi$, where
the left side of the implication connective is always an equality.
For more compact output, our implementation uses a notation that abbreviates
this form of implication. For instance, $\boxm{x=y}\phi$ abbreviates $(x=y)\supset\phi$.
More generally, $\boxm{x_1\!=\!y_1,\cdots,x_n\!=\!y_n}\phi$ abbreviates
$(x_1\!=\!y_1)\supset\cdots\supset(x_n\!=\!y_n)\supset\phi$.
This abbreviation is also used in the screenshot of Figure~\ref{fig:ihaskell}.

The tool can also displaying multiple pairs of distinguishing formulae and displaying
the entire bisimulation tree, which can be useful since the first distinguishing formula generated is not necessarily the most insightful.
A single pair of distinguishing formulae is enough to witness distinguishability
(or, non-bisimilarity) and the entire bisimulation tree is rarely
required to generate a pair of distinguishing formulae. Our Haskell
source code conceptually computes over the structure of the entire
bisimulation tree in order to generate distinguishing formulae.
However, only part of the tree that is needed for the formulae construction
would actually be computed, thanks to Haskell's lazy evaluation,
unless the entire tree is needed elsewhere. Further details of
our algorithm implementation can be found in the technical report
on our initial implementation~\cite{AhnHorTiu17genwit}.

\section{Conclusion} \label{sec:conc}

The main result of this paper is a sound and complete logical characterisation of open bisimilarity for the $\pi$-calculus.
To achieve this result, we introduce modal logic $\mathcal{OM}$, defined in Fig.~\ref{figure:om}.
The soundness of $\mathcal{OM}$ with respect to open bisimilarity, Theorem~\ref{theorem:sound},
is mechanically proven in Abella as explained in Section~\ref{sec:abella}.
The details of the completeness, Theorem~\ref{theorem:complete}, are provided in Section~\ref{sec:complete}.

Intuitionistic modal logic $\OM$ satisfies the following established criteria for an intuitionistic modal logic~\cite{Simpson94}:
\begin{itemize}
\item
Intuitionistic $\mathcal{OM}$ is a conservative extension of intuitionistic logic. Removing modalities, we obtain a standard semantics of intuitionistic modal logic without any new theorems.

\item
Intuitionistic $\OM$ satisfies intuitionistic hereditary. Every operator is closed under an accessible world relation, as given by our Kripke semantics in the Appendix, and also as captured by the notion of respectful substitution in the body of the paper.

\item The law of excluded middle is invalidated. As demonstrated in Examples~\ref{eg:law} and~\ref{eg:elaborate}, the absence of the law of excluded middle is essential for the existence of distinguishing formulae in $\mathcal{OM}$ for processes that are not open bisimilar but are late equivalent.

\item As explored in Corollary~\ref{cor:late}, if we induce the law of excluded middle, we obtain a classical modal logic (characterising late equivalence).

\item In contrast to classical modal logics, diamond and box modalities have independent interpretations, not de Morgan dual to each other.
\end{itemize}

A more direct proof theory for \OM\ is left as an open problem.
A proof system can be used to confirm criteria such as: if $\phi \vee \psi$ has a proof, then either $\phi$ has a proof or $\psi$ has a proof.
A sound and complete proof system would be a step towards addressing the following, more philosophical, criterion for an intuitionistic modal logic~\cite{Simpson94}:
\begin{quote}
There is an intuitionistically comprehensible explanation of the
meaning of the modalities, relative to which IML is sound and complete.
\end{quote}

Previous work on intuitionistic modal logic for program analysis~\cite{Plotkin1986,Steffen94} was motivated by topological interpretations of liveness properties.
The intuitionistic information partial ordering in that work is quite different from in $\OM$, where the intuitionistic information partial ordering is given by the instantiation of inputs. We expect creative use of intuitionistic information partial orderings will lead to further useful intuitionistic modal logics.

The main novelty of this paper is the completeness proof, Proposition~\ref{proposition:non-bisim}, involving an algorithm constructing distinguishing formulae for processes that are not open bisimilar.
To use this algorithm, firstly attempt to prove that two processes are open bisimilar.
If they are not open bisimilar, after a finite number of steps, a distinguishing strategy, according to Def.~\ref{def:non-bisim}, will be discovered.
The strategy can then be used to inductively construct two distinguishing formulae, one biased to each process.
A key feature of the construction is the use of preconditions and diamond for the leading process, e.g., $x = y \yields \diam{\pi}\ttt$,
and box and postconditions for the following process, e.g.,
$\boxm{\pi}\left( x = y \right)$.
Interesting examples involving postconditions are provided in Section~\ref{section:egs}.

The logic $\OM$ is suitable for
formal and automated reasoning.
It has natural encodings
in Abella for mechanised reasoning, used to establish Theorem~\ref{theorem:sound}.
In addition, our distinguishing formulae generation algorithm is implemented in Haskell, as explained in Section~\ref{sec:dfalgo} and a companion report~\cite{AhnHorTiu17genwit}.
We envision that $\OM$ and related intuitionistic modal logics characterising bisimilarity congruences have a role in symbolic model checking.

\section*{Acknowledgments}
We are grateful to Sam Staton for providing an example that helped us discover the completeness proof.
We also appreciate the comments of the anonymous reviewers.
The authors receive support from MOE Tier 2 grant MOE2014-T2-2-076.
The first author receives support from NRF grant 2018R1C1B5046826 funded by Korea government (MSIT).
The second author receives support from Singapore NRF grant NRF2014NCR-NCR001-30.

\vfill
\pagebreak

\bibliographystyle{halpha}
\bibliography{modal}

\pagebreak
\appendix

\section{A Kripke semantics for $\OM$}
\label{sec:kripke}

To provide another explanation for why $\OM$ is intuitionistic,  we provide here a reformulation of the semantics of $\OM$ in Figure~\ref{figure:om} in terms of an intuitionistic Kripke semantics.

In the following, we denote with $\Ecal$ (possibly with subscripts) a finite binary relation between variables. We write $\fv{\Ecal}$ to denote the set of variables occurring in $\Ecal$, and we write $\Ecal^{*}$ to denote the reflexive-symmetric-transitive closure of $\Ecal.$ That is, $\Ecal^{*}$ is an equivalence relation on variables.  
Given a substitution $\sigma$ and a relation $\Ecal$,
we write $\sigma \Vdash \Ecal$ iff 
$x\sigma = y\sigma$ for every $(x,y) \in \Ecal.$

A {\em world} is a triple $(P, h, \Ecal)$ of process $P$, history $h$ and a binary relation $\Ecal$ on variables such that
\begin{itemize}
\item $\fv P \subseteq \fv h$, and $\fv{\Ecal} = \fv{h}$;
\item there exists a substitution $\sigma$ respecting $h$ such that $\sigma \Vdash \Ecal$. 
\end{itemize}
Let us denote with $W$ the set of worlds. Define a relation $\preceq$ on worlds as follows:
\[
(P_1, h_1, \Ecal_1) \preceq (P_2, h_2, \Ecal_2) \quad  \mbox{ iff } \quad P_1 = P_2, ~ h_1 = h_2, ~ \mbox{ and } \Ecal_1^{*} \subseteq \Ecal_2^{*}. 
\]
It is easy to see that $\preceq$ is a partial order on $W$ (reflexive, anti-symmetric and transitive), so the pair $(W ,\preceq)$ forms an intuitionistic Kripke frame. 
A Kripke model for $\OM$ is then a triple $(W, \leq, w)$ where $w \in W.$ In the following, we fix the Kripke frame to $(W,\preceq)$ so it will be implicitly assumed in the definition of the satisfiability relation. Given a world $(h,\Ecal)$ and a substitution $\sigma$, we say that 
$\sigma$ respects $(h,\Ecal)$ iff $\sigma$ respects $h$ and $\sigma \Vdash \Ecal.$

We write $P \VDash^{h,\Ecal} \phi$ when $\phi$ is true in the world $(P, h, \Ecal).$ The complete definition of this satisfaction relation is given in Figure~\ref{figure:om-alt}.
In the figure, the relation $[P \lts{\pi} Q]^{h,\Ecal}$ is defined as 
\[
[P \lts{\pi} Q]^{h,\Ecal} \quad \mbox{iff} \quad \forall \sigma. \mbox{ s.t. $\sigma$ respects $(h,\Ecal)$ and $\bn{\pi} \not \in \fv{\sigma}$, we have  
 $P\sigma \lts{\pi\sigma} Q\sigma.$}
\]
\begin{lem}
\label{lm:respect-monotone}
If $[P_1 \lts{\pi} Q]^{h_1,\Ecal_1}$ and $(P_1,h_1,\Ecal_1) \preceq (P_2,h_2,\Ecal_2)$ then $[P_2 \lts{\pi} Q]^{h_2,\Ecal_2}.$ 
\end{lem}
\begin{proof}
By the definition of $\preceq$, we have $P_1 = P_2$, $h_1 = h_2$ and $\Ecal_1 \subseteq \Ecal_2.$ The lemma then follows
from the fact that if $\sigma$ respects $(h,\Ecal_2)$ then it also respects $(h,\Ecal_1).$
\end{proof}

\begin{lem}
\label{lm:lts-abs}
For every world $(P, h, \Ecal)$,  $\sigma$, $\pi$ and $Q$ such that $\sigma$ respects $(h,\Ecal)$,  $dom(\sigma) \subseteq \fv{h}$, and 
$P\sigma \lts{\pi}  Q$, where $\bn{\pi} \not \in \fv{h} \cup \fv{\sigma}$,
there exists $Q'$ and $\pi'$ such that $\pi = \pi'\sigma$ and $Q = Q'\sigma$ and $[P \lts{\pi'} Q']^{h,\Ecal}.$
\end{lem}
\begin{proof}
By structural induction on the derivation of $P\sigma \lts{\pi} Q.$ We show here an interesting case where match is involved,
e.g., when $P = \match{x=y}R$, and $x\sigma = y\sigma = u$, so $P\sigma = \match{u = u}R\sigma.$
\[
\infer[]
{\match{u=u}R\sigma \lts{\pi} Q}
{R\sigma \lts{\pi} Q} 
\]
By the induction hypothesis, we have $Q'$ and $\pi'$ such that $[R \lts{\pi'} Q']^{h,\Ecal}.$ To show 
$[\match{x = y}R \lts{\pi'} Q']^{h,\Ecal}$ we need to show that for every $\theta$ that respects $(h,\Ecal)$, we have $\match{x\theta = y\theta}R\theta \lts{\pi'\theta} Q'\theta.$
This is constructed as follows:
\[
\infer[]
{\match{x\theta=y\theta}R\theta \lts{\pi'\theta} Q'\theta}
{R\theta\lts{\pi'\theta} Q'\theta} 
\]
The premise follows from $[R \lts{\pi'} Q']^{h,\Ecal}$. It remains to show that this application of the match rule is valid, i.e., that $x\theta = y\theta.$
Since $x\sigma = y\sigma$, and since $(P,h,\Ecal)$ is a world (which means $\fv{P} \subseteq \fv{h} = \fv{\Ecal}$),  and $\sigma$ respects $(h,\Ecal)$, it follows that
$(x,y) \in \Ecal.$ Therefore any $\theta$ respecting $(h,\Ecal)$ would satisfy $x\theta = y\theta.$
\end{proof}

The following results, easy to establish by induction, show that we can translate between the Kripke semantics in Figure~\ref{figure:om-alt} and the semantics in Figure~\ref{figure:om}.
\begin{prop}[Monotonicity] 
\label{prop:monotonicity}
If $(P_1, h_1, \Ecal_1) \preceq (P_2, h_2, \Ecal_2)$ and $P_1 \VDash^{h_1, \Ecal_1} \phi$ then $P_2 \VDash^{h_2, \Ecal_2} \phi.$ 
\end{prop}
\begin{proof}
By induction on the definition of $P_1 \VDash^{h_1,\Ecal_1} \phi$ and Lemma~\ref{lm:respect-monotone}.
\end{proof}

\begin{prop}
\label{prop:om-equiv}
$P \VDash^{h,\Ecal} \phi$ if and only if $P\sigma \models^{h\sigma} \phi\sigma$ for every $\sigma$ that respects $(h,\Ecal).$
\end{prop}
\begin{proof}
This follows from the definition of $\models^h$ and $\VDash^{h,\Ecal}$, and Lemma~\ref{lm:lts-abs}.
\end{proof}

As a consequence of the result above, we obtain the following precise relationship between our two equivalence presentations of the semantics of $\OM$.
\begin{cor}
$P \VDash^{h,id} \phi$ if and only if $P\models^h \phi$, where $id$ is the identity relation on $\fv{h}.$
\end{cor}
\begin{proof}
This is a consequence of Proposition~\ref{prop:om-equiv} and Lemma~\ref{lemma:mono}.
\end{proof}

\begin{figure}
\[
\begin{array}{lcl}
P \VDash^{h,\Ecal} \ttt  &  & \mbox{always holds.} \\
P \VDash^{h,\Ecal} x=y   & \mbox{iff} & (x,y) \in \Ecal \\
P \VDash^{h,\Ecal} \phi_1 \land \phi_2 &\mbox{iff}&
  P \VDash^{h,\Ecal} \phi_1 ~\mbox{and}~ P \VDash^{h,\Ecal} \phi_2.
\\
P \VDash^{h,\Ecal} \phi_1 \lor \phi_1 &\mbox{iff}&
  P \VDash^{h,\Ecal} \phi_1 ~\mbox{or}~ P \VDash^{h,\Ecal} \phi_2.
\\
P \VDash^{h,\Ecal} \phi_1 \yields \phi_1 &\mbox{iff}&
 \forall (P',h',\Ecal') \geq (P,h,\Ecal). ~ 
  P \VDash^{h', \Ecal'} \phi_1 ~ \mbox{ implies } ~ P \VDash^{h', \Ecal'} \phi_2.
\\
P \VDash^{h,\Ecal} \diam{\alpha}\phi &\mbox{iff}&
  \exists\,Q,~   [P \lts{\alpha} Q]^{h,\Ecal} ~\mbox{and}~ Q \VDash^{h,\Ecal} \phi.
\\
P \VDash^{h,\Ecal} \diam{\co{a}(z)}\phi &\mbox{iff}&
  \exists\,Q,~ [P \lts{\co{a}(z)} Q]^{h,\Ecal} ~\mbox{and}~ Q \VDash^{h\cdot z^o, \Ecal_z} \phi.
\\
P \VDash^{h,\Ecal} \diam{{a}(z)}\phi &\mbox{iff}&
  \exists\,Q,~ [P \lts{{a}(z)} Q]^{h,\Ecal} ~\mbox{and}~ Q \VDash^{h\cdot z^i,\Ecal_z} \phi.
\\
P \VDash^{h, \Ecal} \boxm{\alpha}\phi &\mbox{iff}&
  \forall (P',h',\Ecal') \geq (P,h,\Ecal),~
    \forall Q,\, [P' \lts{\alpha} Q]^{h',\Ecal'}  \; \mbox{ implies } \;
    Q \VDash^{h',\Ecal'} \phi.
\\
P \VDash^{h,\Ecal} \boxm{\co{a}(z)}\phi &\mbox{iff}&
   \forall (P',h',\Ecal') \geq (P,h,\Ecal),~ 
   \forall Q, [P' \lts{\co{a}(z)} Q]^{h',\Ecal'} \mbox{ implies } Q \VDash^{h' \cdot z^o, \Ecal'_z} \phi.
\\
 P \VDash^{h,\Ecal} \boxm{{a}(z)}\phi &\mbox{iff}&
   \forall (P',h',\Ecal') \geq (P,h,\Ecal),~ 
  \forall Q, 
    [P' \lts{a(z)} Q]^{h',\Ecal'} \mbox{ implies } Q \VDash^{h' \cdot z^i, \Ecal'_z} \phi.
\end{array}
\]
\caption{An alternative semantics of  \OM. In the figure, $\Ecal_z$ denotes $\Ecal \cup \{(z,z)\}.$ }
\label{figure:om-alt}
\end{figure}

\end{document}